\documentclass[prl,aps,twocolumn,showpacs,showkeys,preprintnumbers,longbibliography,superscriptaddress,amsmath,amssymb,floatfix,footinbib,a4paper,10pt]{revtex4-2}

\usepackage{graphicx}
\usepackage[usenames,dvipsnames]{color}
\usepackage[normalem]{ulem}

\newcommand{\br}{{\bf r}}

\newcommand{\bq}{{\bf q}}
\newcommand{\bV}{{\bf V}}
\newcommand{\bu}{{\bf u}}

\newcommand{\pot}{{\mathbb{W}}}
\newcommand{\corr}{{\xi}}

\newcommand{\eq}[1]{Eq.~(\ref{#1})}

\newcommand{\fref}[1]{Fig.~\ref{#1}}

\newcommand{\rcite}[1]{Ref.~\cite{#1}}

\begin{document}

\title{Granular fluid in an arbitrary external potential: spontaneous
  convection, self-phoresis}

\author{Alvaro Dom\'\i nguez}
\email{\texttt{dominguez@us.es}}
\affiliation{F\'\i sica Te\'orica, Universidad de Sevilla, Apdo.~1065, 
41080 Sevilla, Spain}
\affiliation{Instituto Carlos I de F\'\i sica Te\'orica y Computacional,
  18071 Granada, Spain}

\author{Nagi Khalil}
\email{\texttt{nagi.khalil@uah.es}}
\affiliation{Departamento de F\'isica y Matem\'aticas \& Grupo Interdisciplinar de Sistemas Complejos, Universidad de Alcal\'a, 28805 Alcal\'a de Henares, Spain}

\date{\today}

\begin{abstract}
  The hydrodynamic stationary states of a granular fluid are addressed
  theoretically when subject to energy injection and a
  time--independent, but otherwise arbitrary external potential
  force. When the latter is not too symmetrical in a well defined
  sense, we show that a quiescent stationary state does not exist,
  rather than simply being unstable and, correspondingly, a steady
  convective state emerges spontaneously. We also unveil an unexpected
  connection of this feature with the self-diffusiophoresis of
  catalytically active particles: if an intruder in the granular fluid
  is the source of the potential, it will self-propel according to a
  recently proposed mechanism that lies beyond linear response theory,
  and that highlights the role of the intrinsic nonequilibrium nature
  of the state of the granular bath.  In both scenarios, a
  state--dependent characteristic length of the granular fluid is
  identified which sets the scale at which the induced flow is the
  largest.
\end{abstract}



\maketitle

The research of the out-of-equilibrium behavior of many-body systems
has experienced a sustained interest for over half a century,
particularly as a means of advancing in the comprehension of
complexity \cite{deMa84,SSG07,BBGR17,Walk17}. For long, the
experiments and the corresponding models addressing the
nonequilibrium phenomenology focused on macroscopic external
gradients in order to drive the system away from equilibrium
\cite{deMa84}, meant to describe the effect of the interaction between
the system and the outer world.  However, the last years have
witnessed a shift in interest towards sources of nonequilibrium that
appear at the scale of the putative constituents (``microscopic
level''): some notorious instances of these ``active
systems'' are the granular fluids (where energy conservation is
violated due to inelastic particle--particle collisions and local
injection) \cite{Camp90,JNB96,Gold03,BTE05}, the collections of
self-propelled particles intended to model bird flocks, fish schools,
and the like (where, in addition, momentum conservation is violated
due to self-propulsion and non-reciprocal interactions)
\cite{Rama10,ViZa12,ZoSt16,GSKS24}, and the colloids composed of
chemically active particles as realizations of microswimmers (where,
in addition, mass conservation is violated by the catalytic action
of the particles on the solvent) \cite{PKOS04,FAMO05,GLA05,EbHo10,MoPo17}.

A particularly interesting question concerns the interplay between the
intrinsic dynamics of the active system and an externally imposed
conservative force field acting on its constituents. When the
postulates of thermodynamics hold, the time--independent force field
does not prevent the system from reaching an equilibrium state, albeit
spatially inhomogeneous. And indeed, the basic result underlying
the Density Functional Theory, see, e.g., \rcite{Evan79}, is the
existence of a one-to-one mapping between the external potential and
the particle distribution. A similar result for out-of-equilibrium is
not known and, focusing on granular fluids, the research has usually
addressed very symmetrical configurations of the external force
fields, like a spatially homogeneous force (gravity), usually aligned
with the container walls, see, e.g.,
\cite{BRM01,WHP01,TaVi02,WRP13,PGVP16,Khal16,WLMP18,VPPG19}; the role
of symmetry breaking in a granular fluid has received attention only
recently, see, e.g., \cite{MPC20,ZhPo21,PlPu22}. We here study the
solutions of the hydrodynamic equations for a fluidized granular
system in an arbitrary external conservative force field, and derive
insightful results related specifically to the force symmetries.

\textit{Theoretical model.---} We consider a fluid of inelastic,
identical particles of mass $m$, that are subject to an external
source of energy and to an external field of force characterized by
the potential energy field $\pot(\br)$. The macroscopic state is
described by the hydrodynamic fields: a particle number density
$n(\br, t)$, a flow velocity $\bu(\br, t)$, and a kinetic or granular
temperature \cite{PSV17} $T(\br, t)$. The evolution follows a set of
hydrodynamic equations expressing the balance of mass, momentum, and
energy, respectively, in spatial dimension $D$ (here,
$d/dt := \partial/\partial t + \bu\cdot\nabla$ is the Lagragian time
derivative):
\begin{equation}
  \label{eq:cont}
  \frac{d n}{d t} = - n \nabla\cdot\bu ,
\end{equation}
\begin{equation}
  \label{eq:vel}
  m n \frac{d \bu}{d t}
  = - \nabla p - n \nabla\pot + \nabla\cdot \mathsf{\Sigma} ,
\end{equation}
\begin{equation}
  \label{eq:temp}
  \frac{D}{2} n \frac{d T}{d t}
  = - \frac{D}{2} n\, G - p \nabla\cdot\bu
  +   \mathsf{\Sigma} : (\nabla\bu)
  - \nabla\cdot\bq .
\end{equation}
Here we have introduced the pressure $p$, the viscous stress tensor
$\mathsf{\Sigma} := \eta \left[ \nabla \bu + (\nabla \bu)^\dagger -
  (2/D) \mathsf{I}\, \nabla\cdot\bu \right]$, the heat current density
$\bq := - (\kappa \nabla T + \mu \nabla n)$, and the energy source
term $G = \zeta T - Q$ that incorporates both the dissipation by
inelastic collisions ($\zeta T$), and the bulk energy injection $Q$
from an external source, e.g., a Gaussian thermostat
\cite{WiMa96,PeOh98}, a fixed addition of kinetic energy at each grain
collision \cite{BRS13}, the fast vibrations of a plate over which a
monolayer of grains resides \cite{OlUr98,MVPR05}, or a strong air
current through the granular medium \cite{LBLG00}.  The equations
feature the transport coefficients: the cooling rate $\zeta$ due to
the inelasticity, the shear viscosity $\eta$, the thermal conductivity
$\kappa$, and the thermal diffusivity $\mu$ that accounts for the heat
flow driven by density gradients, as a modification of Fourier's law
\cite{BDKS98}. Consistently with the assumption that the hydrodynamic
fields $n,\bu, T$ provide the complete macroscopic description, the
functions $p, G$ and the transport coefficients are assumed to depend
on the local values of the scalar fields $n$ and $T$, so that
Eqs.~(\ref{eq:cont}--\ref{eq:temp}) form a closed set of equations.
The elastic limit of the equations corresponds to $\zeta=0$, $G=0$,
$\mu=0$.

\textit{Quiescent states.---} Since the external field $\pot(\br)$ is
time independent, there could conceivably exist stationary
($\partial_t\equiv 0$), quiescent ($\bu\equiv 0$) states, 
characterized by spatially inhomogeneous profiles $n(\br)$ and
$T(\br)$ in number density and granular temperature. The continuity
equation~(\ref{eq:cont}) is satisfied automatically, and the equation
for momentum balance~(\ref{eq:vel}) reduces to hydrostatic
equilibrium,
\begin{equation}
  \label{eq:stat1}
  \nabla p 
  = p_n \nabla n + p_T \nabla T = - n \nabla\pot ,
\end{equation}
where, in the following, we assume the generic case
$p_n := \partial p/\partial n \neq 0$,
$p_T := \partial p/\partial T \neq 0$, i.e., far from the dissipative
analog of a possible ``phase transition''. The curl of this
equation leads to the constraints \cite{SuppInfo}
\begin{equation}
  \label{eq:nTxphi}
  \nabla n \times \nabla \pot = 0 ,
  \quad
  \nabla T \times \nabla \pot = 0 ,
  \quad
  \nabla n \times \nabla T = 0 ,
\end{equation}
that is, the isopycnic, the isothermal, and the equipotential
surfaces, respectively, coincide (this includes trivially the case
that some gradient vanishes identically, e.g., an equilibrium state,
for which $\nabla T \equiv 0$). These geometrical constraints imply
\cite{SuppInfo} that the $\br$-dependence of the fields $n$, $T$ can
be expressed completely via the external potential: in a spatial
domain where the latter is strictly monotonic ($\nabla\pot \neq 0$),
there exist certain functions $\nu(\pot)$, $\tau(\pot)$ such that it
is possible to write $n(\br) = \nu (\pot(\br))$ and
$T(\br) = \tau (\pot(\br))$. But then, upon defining the functions
$\alpha(\pot) := (D/2)\, \nu \, G(\nu, \tau)$ and
$\beta(\pot) := \kappa (\nu, \tau)\, d\tau/d\pot
+ \mu (\nu, \tau)\, d\nu/d\pot$, 
\eq{eq:temp} for energy balance,
\begin{equation}
  \label{eq:stat2}
  \frac{D}{2} n G = -\nabla\cdot\bq = \nabla\cdot\left( \kappa \nabla T
    + \mu \nabla n \right) ,
\end{equation}
can be written equivalently as
\begin{equation}
  \label{eq:stat3}
  \alpha(\pot) =   \nabla \cdot\left[ \beta(\pot) \nabla\pot \right]
  = \beta(\pot) \nabla^2\pot + \frac{d \beta}{d\pot} |\nabla\pot|^2 .
\end{equation}
This latter expression represents a consistency constraint on the
external potential: there must exist a linear relationship
between the fields $|\nabla\pot|^2$ and $\nabla^2\pot$ on each
equipotential surface. This is not the case unless the potential field
has a high symmetry, e.g., gravity
($\pot=m g z\,\Rightarrow\, |\nabla\pot|^2=(mg)^2, \nabla^2\pot=0$),
or an isotropic harmonic well
($\pot=k r^2\,\Rightarrow\, |\nabla\pot|^2 =4 k \pot,
\nabla^2\pot=2Dk$); already an anisotropic harmonic potential violates
the constraint due to the misalignment of the iso--surfaces of the
fields $\pot=\sum_{i} k_i x_i^2$,
$|\nabla\pot|^2 =4 \sum_i (k_i x_i)^2$, and
$\nabla^2\pot=2\sum_i k_i$. In summary, there cannot exist a
stationary, quiescent state for an arbitrary external potential. This
statement is stronger than the instability of an existing stationary,
quiescent state, which is a rather common scenario observed for
granular fluids \cite{Haff83,RLV20}.

Another interesting consequence follows by applying this argument in
the elastic limit ($G\equiv 0 \Rightarrow \alpha\equiv 0$), because
then there always exists a state of thermodynamic equilibrium
(stationary, quiescent, and with uniform temperature
$\Rightarrow d\tau/d\pot=0$) for \emph{any form} of the potential
field $\pot(\br)$. Therefore, the consistency
constraint~(\ref{eq:stat3}), which becomes
$\nabla\cdot[\mu(\pot) (d\nu/d\pot) \nabla \pot]=0$, must be always
satisfied, and this occurs only if $\mu$ vanishes because
\eq{eq:stat1} prevents $d\nu/d\pot=0$. This provides further insight
into why the coefficient $\mu$ in Fourier's law must vanish for
elastic fluids, in spite of not being forbidden by overt symmetry
considerations (as evidenced by its appearance when \eq{eq:temp} is
derived from a kinetic model \cite{JeSa83,SeGo98,BDKS98,Duft00}), and
complements the proof based on entropic arguments \cite{LaLi6}.

\begin{figure}
  \hfill
  \begin{tabular}[c]{cc}
    & \includegraphics[width=.35\textwidth]{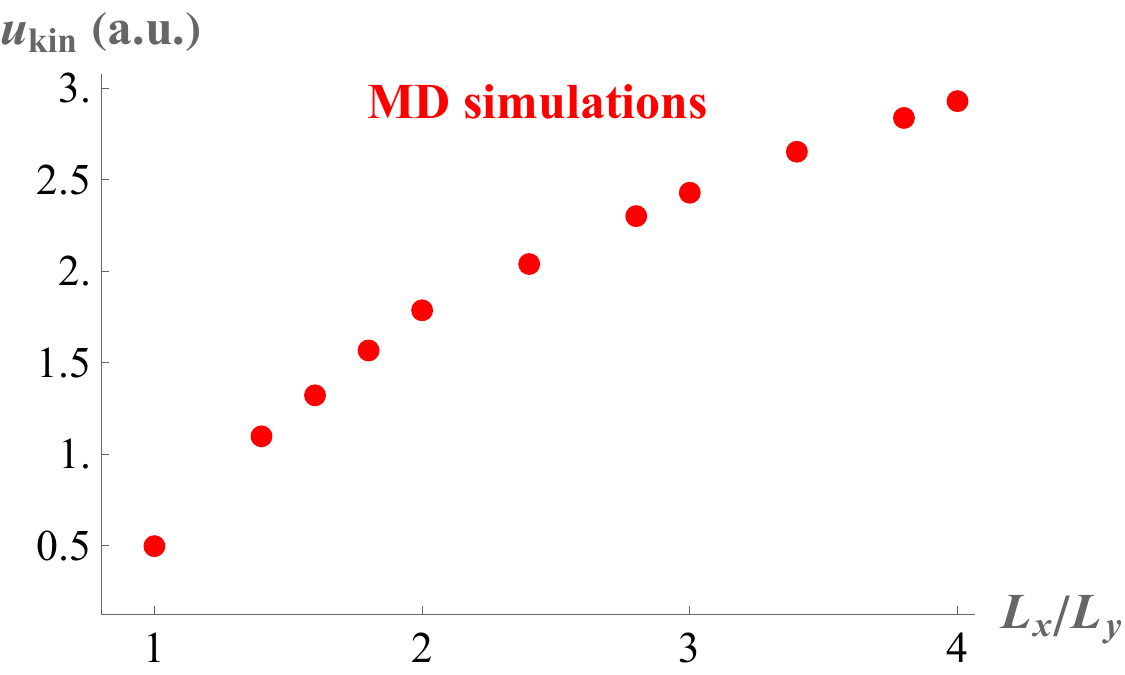}
    \\
    \\
    \begin{tabular}[c]{c}
      $|\bu|/u_\mathrm{kin}$
      \\
      \includegraphics[width=.05\textwidth]{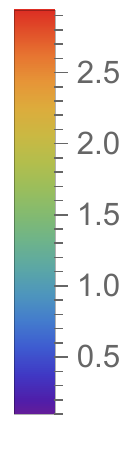}
    \end{tabular}
    &
      \begin{tabular}[c]{c}
        MD simulations, $L_x/L_y=4.0$, $N=100$
        \\
        \includegraphics[width=.35\textwidth]{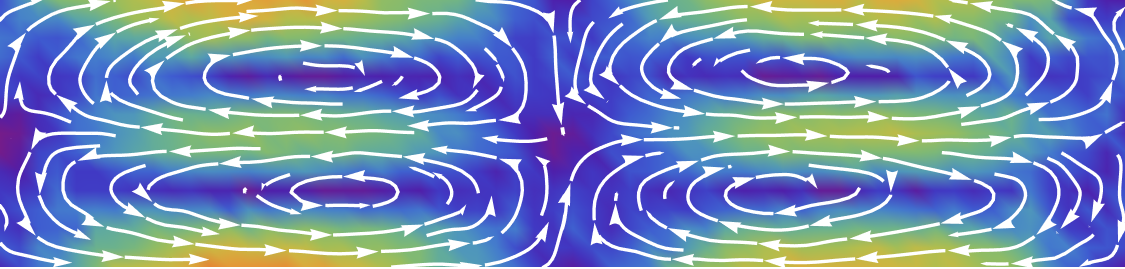}
        \\
        \\
        perturbative theory, $L_x/L_y=4.0$
        \\
        \includegraphics[width=.35\textwidth]{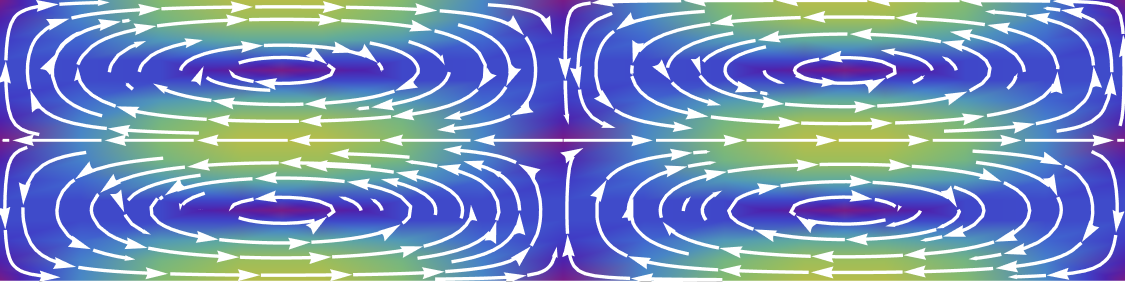}
      \end{tabular}
  \end{tabular}
  \hspace*{\fill}
  \caption{Results for the flow induced in a 2D
    granular fluid by a potential
    $\pot(\br) = -\cos (2\pi x/L_x) - \cos(2\pi y/L_y)$ in a
    rectangular box of side--lengths $L_x$ and $L_y$ with periodic
    boundary conditions. The strength of the flow is characterized by
    the velocity scale $u_\mathrm{kin}:= \sqrt{2 K/m N}$, where $K$ is
    the total kinetic energy of the flow and $N$ is the number of
    grains.  The upper plot shows the growth of $u_\mathrm{kin}$ (in arbitrary units)
    with the aspect ratio $L_x/L_y$ extracted from Molecular Dynamics,
    consistently with the perturbative prediction \cite{SuppInfo}.
    The lower pannel compares the flow in simulations with the
    perturbative computation for a particular value of the aspect
    ratio $L_x/L_y$ (the same conclusions hold for other values, see
    \rcite{SuppInfo}). The white arrows are the streamlines, the
    colored background encodes the modulus of the velocity field
    normalized by $u_\mathrm{kin}$. The estimated Reynolds and Mach
    numbers for the simulations are $\sim 10^{-3}$, so that
    Eqs.~(\ref{eq:divu},\ref{eq:flowlin}) for the flow field are
    likely a good approximation. The variations in particle number
    density across the domain amount however to up to $70\%$ of the
    average density, which casts doubt on the validity of
    Eqs.~(\ref{eq:plin},\ref{eq:heatlin}) and the form of the forcing
    term in \eq{eq:flowlin}; nevertheless, the perturbative theory
    captures the relevant features of the measured flow.
    \label{fig:sim}
  }
\end{figure}

\textit{Convection.---} In the absence of stationary states that are
quiescent, one may look for steady states of non-vanishing flow, i.e.,
solutions to Eqs.~(\ref{eq:cont}--\ref{eq:temp}) with $\partial_t=0$
but $\bu\not\equiv 0$. Qualitative understanding can be gained by
addressing the effect of a weak external potential
perturbatively. Therefore, we assume that, when $\nabla\pot\equiv 0$,
there exists a stationary and homogeneous quiescent state,
$n(\br, t)=n^{(0)}, T(\br, t)=T^{(0)}, \bu(\br, t)=0$, determined by
the condition $G(n^{(0)}, T^{(0)})=0$ and which is linearly stable, as
is actually the case when different models of energy injection are considered
\cite{WiMa96,PeOh98,BRS13,OlUr98,MVPR05}.  A perturbative calculation
\cite{SuppInfo} then provides the equations to leading order in
$\pot$, which are linear for the density and the temperature, and
quadratic for the velocity. For this purpose, it is convenient to
define the \textit{``heat potential''} field,
\begin{equation}
  \label{eq:rho}
  \rho(\br) := n(\br) - n^{(0)}+ \frac{n^{(0)}\kappa^{(0)}}{p_n \kappa^{(0)} - p_T
    \mu^{(0)}}  \pot(\br), 
\end{equation}
so called because it is related to the heat flux at this order of
approximation as
$\bq = -\nabla (\kappa^{(0)} T + \mu^{(0)} n)=p_T^{-1} (p_n
\kappa^{(0)} - p_T \mu^{(0)}) \nabla \rho$. It is also useful to
introduce two parameters associated to the reference state of the
system that contain all the explicit dependence on the inelasticity: a
characteristic length
\begin{equation}
  \label{eq:corr}
  \corr := \sqrt{\frac{2}{D n^{(0)}} \, \frac{p_n \kappa^{(0)} - p_T \mu^{(0)}}{p_n G_T
    - p_T G_n}},
\end{equation}
and the factor
\begin{equation}
  \label{eq:gamma}
  \gamma := \frac{n^{(0)} p_T}{p_n \kappa^{(0)} - p_T \mu^{(0)}}\;
    \frac{\kappa^{(0)} G_n - \mu^{(0)} G_T}{p_n G_T
      - p_T G_n} ,
\end{equation}
whereby the transport coefficients
$\eta^{(0)}, \kappa^{(0)}, \mu^{(0)}$ and the derivatives
$p_n, p_T, G_n, G_T$ are evaluated at the homogeneous, quiescent
state. In terms of these quantities, the conditions of hydrostatic
equilibrium and energy balance yield, after linearizing
Eqs.~(\ref{eq:stat1}) and (\ref{eq:stat2}) respectively,
\begin{equation}
  \label{eq:plin}
  \nabla p = p_n \nabla n  + p_T \nabla T = - n^{(0)} \nabla \pot ,
\end{equation}
\begin{equation}
  \label{eq:heatlin}
  \corr^2 \nabla^2 \rho - \rho = \gamma \pot ,
\end{equation}
complemented by the equations of incompressible, creeping flow
following from Eqs.~(\ref{eq:cont}) and (\ref{eq:vel}),
\begin{equation}
  \label{eq:divu}
  \nabla\cdot\bu=0 ,
\end{equation}
\begin{equation}
  \label{eq:flowlin}
  \eta^{(0)} \nabla^2 (\nabla\times\bu) = - \frac{1}{n^{(0)}}
  \nabla \times
  \left( n \nabla p \right)
  = \nabla \rho \times \nabla\pot .
\end{equation}
The length $\corr$ is a real quantity because linear stability of the
reference state requires $G_T>0, p_n G_T - p_T G_n>0$ \cite{BRS13}
and, if the inelasticity is not too large,
$p_n \kappa^{(0)} - p_T \mu^{(0)}>0$ in order for the heat to flow
against the temperature gradient \cite{SuppInfo}. This length follows
from the interplay between the heat flux $\bq$ and the energy source
term $G$ in \eq{eq:stat2}. The sign of $\gamma$, on the contrary, is
undefined and depends critically on the degree of inelasticity through
the sign of the coefficient $\mu^{(0)}$
\cite{GaDu99a,BrCu01,Luts05}. The elastic limit ($G\to 0$ and
$\mu^{(0)}\to 0$) is represented by the double limit
$\corr\sim G^{-1/2}\to\infty$, $\gamma \corr^{-2} \sim G \to 0$,
$\gamma$ finite, which leads naturally to the expected equilibrium
state: $\nabla^2\rho=0 \Rightarrow\rho=0\Rightarrow$ no heat flux, no
flow, and an isothermal, barometric profile $n_\mathrm{eq}(\br)$
determined by the field $\pot(\br)$ \cite{SuppInfo}.

One can derive some insightful consequences from
Eqs.~(\ref{eq:plin}--\ref{eq:flowlin}). The first two equations
provide the inhomogeneous profiles of number density and
temperature. They are in turn responsible for driving a flow according
to the forced Stokes \eq{eq:flowlin}, which describes how the lack of
sufficient symmetry in the potential $\pot(\br)$, which leads to the
violation of the constraints~(\ref{eq:nTxphi}), is reflected in the
generation of vorticity by \emph{baroclinity} \cite{Trit88}, i.e., due
to the misalignment of the gradients in density and pressure. This
flow is thus quadratic in the perturbation $\pot$, and the theory does
address a nonlinear effect.  This mechanism is distinctly different
from the more frequent scenario of a Rayleigh--B\'enard instability
(also observed in a fluidized granular system \cite{RRC00}): although
baroclinity still plays a role, no external temperature gradient is
imposed, and convection is spontaneous because the transition
threshold vanishes \cite{HaWa77}.

The change of variable $\hat{\rho} = \rho + \gamma \pot$ does not
affect baroclinity but \eq{eq:heatlin} becomes
$\corr^2 \nabla^2\hat{\rho} -\hat{\rho} = \corr^2 \gamma \nabla^2
\pot$. Therefore, when the potential is a harmonic function
($\nabla^2 \pot =0 $) no flow is induced; this is a specific example,
at this level of approximation, of the notion that $\pot$ should not
be too symmetric. One can also extract the asymptotic scaling with the
length $\corr$ at fixed $\gamma$ and argue that the amplitude of the
flow field will not change monotonically with $\corr$. When
$\corr\to 0$, the heat potential is screened on scales much shorter
than any characteristic length associated to $\pot(\br)$ and the field
$\hat{\rho}$ is full determined by a completely local relationship,
$\hat{\rho} \approx - \corr^{2} \gamma \nabla^2 \pot$;
correspondingly, the baroclinity $\nabla\hat{\rho}\times\nabla \pot$
is suppressed as $\corr^2$ \footnote{Barring the case that
  $\nabla^2\pot$ vanishes identically. A notorious instance of this
  exception is provided by self-phoresis of an active particle, see
  the discussion of Fig.~E in the Supplemental Material.}. In the
opposite limit, $\corr\to\infty$, the behavior is effectively
quasi--elastic: the screening term can be dropped in the Helmholtz
equation~(\ref{eq:heatlin}), so that
$\nabla^2\rho \approx \corr^{-2} \gamma \pot$ and the baroclinity is
suppressed as $\corr^{-2}$. Therefore, the strength of the flow
induced by $\pot$ will expectedly be maximal in a state with an
intermediate value of the scale $\corr$.

To illustrate the qualitative picture that follows from the
constraints~(\ref{eq:nTxphi}) and from
Eqs.~(\ref{eq:plin}--\ref{eq:flowlin}), \fref{fig:sim} shows results
from numerical simulations and from the perturbative theory for a
particular configuration.

\begin{figure}[t]
    \centering
    \hfill\includegraphics[width=.35\textwidth]{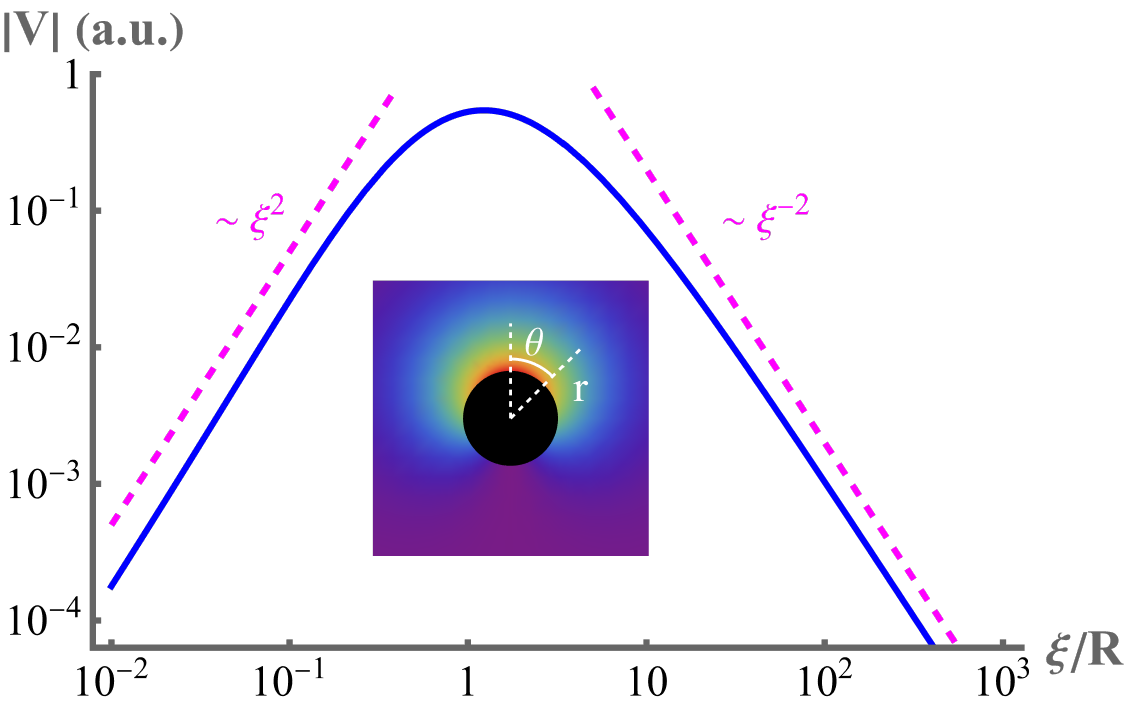}
    \hspace*{\fill}
    \caption{Self-phoretic translation velocity $|\bV|$ (in arbitrary
      units) of a spherical intruder of radius $R$ as a function of
      the ratio $\corr/R$. The intruder exerts a potential on the
      granular fluid bath of the form
      $\pot(\br)= \mathrm{e}^{-r/R} \left( 1 + \cos\theta \right)$,
      shown in the inset as a heat map. The intruder also imposes a
      no-slip boundary condition on the flow at its surface. The
      dashed lines are the expected asymptotic behaviors.}
  \label{fig:selfphoresis}
\end{figure}

\textit{Self-phoresis.---} Consider now the alternative scenario of a
potential field $\pot(\br)$ sourced by a freely moving object (an
intruder) immersed in the granular bath \footnote{Perturbatively in a
  weak potential, one can neglect the time--dependence of $\pot$
  induced by the motion of the intruder to leading order
  \cite{SuppInfo}.}. The nontrivial prediction \cite{SuppInfo}
following from the perturbative
theory~(\ref{eq:plin}--\ref{eq:flowlin}) is that the induced grain
flow will also lead to a directed motion (translation and rotation) of
the intruder: it effectively becomes an \emph{autophoretic swimmer}
because it self-propels while the combined system ``intruder +
granular bath'' is mechanically isolated (i.e., no net external force
or torque, setting this phenomenon fundamentally apart from the case
of external particle drag or fluid stir). More specifically, assuming
for simplicity no-slip boundary conditions for the flow field
$\bu(\br)$ on the surface of the intruder and elastic collisions of
the grains with it, one can apply the Lorentz reciprocal theorem (see,
e.g., \rcite{HaBr73,KiKa91,Teub82,DPRD20}) to the flow
equations~(\ref{eq:divu}, \ref{eq:flowlin}) and find a self-phoretic
velocity of translation as \cite{DoPo24b,DPD25,SuppInfo}
\begin{equation}
  \label{eq:phorV}
  \bV = \int\limits_\mathrm{bath} d^3\br\; \mathsf{M}^{(V)}(\br) \cdot
  \left[ \nabla \pot(\br) \times \nabla \rho(\br) \right] ,
\end{equation}
with a tensor field $\mathsf{M}^{(V)}(\br)$ determined completely by
the geometrical shape of the intruder; a similar result holds for the
rotational velocity but with a different tensor field. This
expression shows explicitly how self-phoresis is driven by the
gradient misalignment.

Quite unexpectedly, this prediction can be framed within the
correlation--induced mechanism that has been recently identified for
the self-diffusiophoresis of a catalytically active particle
\cite{DPRD20,DoPo22}, i.e., the self-propulsion of a particle when it
induces a concentration gradient in the surrounding fluid
solution. The mathematical model shares Eqs.~(\ref{eq:divu})
and~(\ref{eq:flowlin}) for the flow, but \eq{eq:heatlin} is replaced
by the same equation for the solute concentration with the role of
$\pot$ played by the solute chemical potential, the latter being in
turn sourced by the catalytic activity --- see table~I in
\rcite{SuppInfo}; actually, the mathematical model is slightly more
involved than Eqs.~(\ref{eq:heatlin}--\ref{eq:flowlin}). Common to
these two scenarios (granular bath or catalytic activity,
respectively) are that the hydrodynamic flow is driven by a gradient
misalignment and, most indicative, that the misaligned gradients are
generated by the same source (either the potential $\pot$ or the
catalytic activity, respectively). These two features have a relevant
observational impact. First, the phoretic velocities depend
quadratically on that source, in clear contrast with the
linear--response prediction of the ``classic'' mechanism
\cite{DSZK47,DSZK93,Ande89}, so that the correlation--induced mechanism
disproves the paradigm that ``self-phoresis is just normal phoresis
but in a self-induced gradient''. And second, the velocities exhibit a
significant dependence on a length scale that is exclusively
associated to the fluid medium and unrelated to the phoretic particle,
namely, the scale $\corr$ in \eq{eq:heatlin} that parallels the
solute--solute correlation length in the catalytic--activity scenario.

One can learn further by borrowing the analytical results of the
catalytic--activity scenario for the case of a spherical intruder in
an unbounded bath \cite{SuppInfo}. For instance, one obtains a set of
``selection rules'' \cite{DPRD20} on the spherical harmonic expansion
of $\pot(\br)$ that quantify precisely the notion that this field
should not be too symmetric in order to yield a nonvanishing phoretic
velocity. But one also finds insightful differences: the phoretic
angular velocity is non zero in general, and the translational
velocity does not change monotonously with $\corr$: see
\fref{fig:selfphoresis}, which shows the predicted asymptotic
behaviors \footnote{It is illuminating to compare with Fig.~E in the
  Supplemental Material.}. 

\textit{Conclusions.---} The customary use of very simplified
configurations hides the relevant role of the asymmetry of an imposed
external potential, and has instead directed the focus onto the
relevance of boundary conditions that describe energy exchange, e.g.,
through heat flow or inelasticity
\cite{WHP01,TaVi02,WRP13,PGVP16,VPPG19}.  This has eventually led to
the conclusion that an external force (e.g., gravity) is unnecessary
to generate convection \cite{WLMP18,RLV20}. But this conclusion is of
limited scope and not distinctive for a granular fluid because this
kind of ``thermal'' boundaries are also enough to drive an elastic
fluid out of equilibrium.

By addressing, however, a less symmetric conservative force field, one
highlights a specific, fundamental feature brought about by the
intrinsically dissipative granular dynamics, namely the unavoidably
simultaneous appearance of both density and temperature gradients.
One must instead acknowledge that ``thermal'' boundary conditions are
unnecessary in order to drive convective flows: an external potential
suffices, in clear contrast to the case of elastic fluids. The
conceptual difference was exemplified with \fref{fig:sim}: it is not
that a no-flow state becomes unstable (akin to the scenario of a
Rayleigh--B\'enard instability) when the domain aspect ratio departs
from unity, but rather that this state \emph{ceases to exist}. One can
thus view the present work as a first step in the classification of
the solutions to Eqs.~(\ref{eq:cont}--\ref{eq:temp}) as function of
the external potential.

When the conservative force that induces the flow is generated by a
free object (intruder) within the granular fluid, a directed motion of
the intruder is predicted, which therefore qualifies as a
self-phoretic particle (a ``swimmer''). We unveil an unexpected
connection with the recently identified correlation--induced mechanism
of self-diffusiophoresis of catalytically active particles \cite
{DPRD20,DoPo22}. In spite of the formal similarity of the respective
mathematical descriptions, there are, however, three noteworthy and
closely related differences: first, there arise density and
temperature gradients in the granular bath, so that the phenomenon
cannot be properly qualified either as diffusiophoresis or as
thermophoresis: it rather seems to be a hybrid case. Second, the
granular bath is intrinsically out of equilibrium and the intruder
plays no role at that, which is the complementary scenario of a
``normal'' (elastic) fluid driven out of equilibrium by an active
intruder. This feature alters the role played by the intruder--bath
interaction potential $\pot(\br)$: from determining the ``classic''
(linear) response to gradients \cite{Ande89} in the
catalytic--activity scenario, to becoming also the source of those
gradients in the granular bath. And finally, although both instances
of self-phoresis involve a characteristic fluid length $\corr$ in a
highly relevant manner, the natural limits to be considered are
opposite, respectively $\corr\to 0$ in the catalytic--activity case
(leading to the thin--layer and lubrication approximations
\cite{DoPo24a}), and $\corr\to\infty$ in a granular fluid
(corresponding to the elastic limit). In summary, an additional
scenario has been identified to address the fundamentals of
self-phoresis from a fresh perspective.

\begin{acknowledgments}
Financial support is acknowledged through grants ProyExcel\_00505
funded by Junta de Andaluc{\'i}a, and PID2021-126348NB-I00 and
PID2023-151960NA-I00 funded by MCIN/AEI/10.13039/501100011033 and by
``ERDF A way of making Europe''.
\end{acknowledgments}

\end{document}


\title{Granular fluid in an arbitrary external potential: spontaneous
  convection, self-phoresis
  \\
  \texttt{Supplemental Material}\footnote{Here, equations and figures
    of the main text are referenced by their respective
    numbers. Equations appearing only in the Supplemental Material
    are cited by a combination of the section number (in Roman
    numerals) and the equation number within the section (in Arabic
    numerals). Figures appearing only in the Supplemental Material
    are referenced with a capital letter.}  }

\author{Alvaro Dom\'\i nguez}
\email{\texttt{dominguez@us.es}}
\affiliation{F\'\i sica Te\'orica, Universidad de Sevilla, Apdo.~1065, 
41080 Sevilla, Spain}
\affiliation{Instituto Carlos I de F\'\i sica Te\'orica y Computacional,
  18071 Granada, Spain}

\author{Nagi Khalil} \email{\texttt{nagi.khalil@uah.es}}
\affiliation{Departamento de F\'isica y Matem\'aticas \& Grupo
  Interdisciplinar de Sistemas Complejos, Universidad de Alcal\'a,
  28805 Alcal\'a de Henares, Spain}


\maketitle

\tableofcontents

\newpage

\section{The hydrodynamic equations}
\label{sec:hydroEqs}

A frequently employed model for the hydrodynamic evolution of a
$D$--dimensional granular fluid is provided by
Eqs.~(\ref{eq:cont}--\ref{eq:temp}), namely
\begin{subequations}
  \label{eq:hydrofull}
  \begin{equation}
    \label{eq:contfull}
    \frac{\partial n}{\partial t} = - \nabla\cdot (n \bu) ,
  \end{equation}
  \begin{equation}
    \label{eq:velfull}
    m n 
    \left[ 
      \frac{\partial \bu}{\partial t} + (\bu\cdot\nabla)\bu
    \right]
    = - \nabla p(n,T) - n \nabla\pot + \nabla\cdot \mathsf{\Sigma} ,
  \end{equation}
  \begin{equation}
    \label{eq:tempfull}
    \frac{D}{2} n 
    \left[ \frac{\partial T}{\partial t} +
      (\bu\cdot\nabla)T \right]
    = - \frac{D}{2} n\, G(n,T) - p(n,T) \nabla\cdot\bu
    +   \mathsf{\Sigma} : (\nabla\bu)
    - \nabla\cdot\bq ,
  \end{equation}
  \begin{equation}
    \label{eq:stressfull}
    \mathsf{\Sigma} := \eta(n,T) \left[ \nabla \bu +
      (\nabla \bu)^\dagger - \frac{2}{D} \mathsf{I} \;
      \nabla\cdot\bu \right] ,
  \end{equation}
  \begin{equation}
    \label{eq:heatfull}
    \bq := - \left[ \kappa(n,T) \nabla T + \mu(n,T) \nabla n \right],
  \end{equation}
  \begin{equation}
    \label{eq:Gfull}
    G(n,T) := T \zeta(n,T) - Q(n,T) .
  \end{equation}
\end{subequations}
These equations can be postulated at a phenomenological level
\cite{Haff83} as a modification of those for a normal fluid, so that
they account for the effect of the inelasticity of the collisions
between particles. A more rigorous derivation, which yields terms in
the equations overlooked by the phenomenological approach like the
$\mu$--term in the heat flux, relies on an underlying kinetic model,
e.g., a Boltzmann--like equation in the dilute limit
\cite{JeSa83,SeGo98,BDKS98,Duft00,BBMG15}. More specifically, the
hydrodynamic equations can be derived in the form of a gradient
expansion, assuming that the hydrodynamic fields $n$, $\bu$, $T$, and
the external potential $\pot$ vary slowly over the relevant
microscopic scales, e.g., the mean free path in a dilute system.

The equations~(\ref{eq:cont}--\ref{eq:temp}) result from expanding up
to second order in the gradients. In addition to the terms quoted in
those equations, the procedure gives also a term
$\propto \nabla\cdot \bu$ of volume viscosity in the viscous stress
tensor $\mathsf{\Sigma}$ which can be interpreted as a dependence in
the pressure $p$ additional to the dependence on $n$ and $T$, as well
as a linear dependence of the scalar coefficient $G$ on the scalar
combinations
$\nabla\cdot\bu, (\nabla\bu):(\nabla\bu),
(\nabla\bu):(\nabla\bu)^\dagger, \nabla^2 n, \nabla^2 T, |\nabla n|^2,
|\nabla T|^2, (\nabla n)\cdot(\nabla T)$, on top of the dependence on
$n$ and $T$. However, we have neglected these additional dependences
for simplicity because they do not alter the conclusions and, for not
too large inelasticity, are quantitatively small compared to
the retained terms \cite{SeGo98,BDKS98}.

\section{The constraints on the fields}
\label{sec:constraints-fields}

The constraints~(\ref{eq:nTxphi}) follow from taking the curl of the
hydrostatic equilibrium condition~(\ref{eq:stat1}). First, it holds
\begin{equation}
  \nabla\times \nabla p = 0 
  \qquad\Rightarrow \qquad \nabla n\times \nabla \pot=0 
\end{equation}
straightforwardly. When this result is now applied to evaluate the cross
product of \eq{eq:stat1} with $\nabla\pot$, one gets
\begin{equation}
  0 = - n \nabla\pot\times\nabla\pot
  = p_n \overbrace{\nabla\pot\times\nabla n}^{0} +  p_T \nabla\pot\times\nabla T  .
\end{equation}
Likewise, the cross product with $\nabla n$ renders
\begin{equation}
  - n \overbrace{\nabla n\times\nabla\pot}^{0}
  =  p_T \nabla n\times\nabla T .
\end{equation}

The constraints~(\ref{eq:nTxphi}) mean geometrically that the normal
vector to the isopycnic surfaces $n(\br)=\mathrm{constant}$, to the
isothermal surfaces $T(\br)=\mathrm{constant}$, and to the
equipotentials $\pot(\br)=\mathrm{constant}$ are parallel to each
other \emph{at all points}. But this implies that those surfaces must
overlap completely: consider, for instance, the condition of
parallelism of the isopycnic and the equipotential, expressed as
\begin{equation}
  \nabla n(\br) = F(\br) \nabla\pot(\br),
  \qquad
  \textrm{where $F(\br)$ is a scaling factor.}
\end{equation}
Take any path $\mathcal{C}$ lying on an arbitrary equipotential and
connecting two points $\br_A$, $\br_B$. The change in density along
this path can be written as the line integral
\begin{equation}
  n(\br_B) - n(\br_A) = \int_\mathcal{C} d\boldsymbol{\ell}\cdot\nabla
  n(\br)
  = \int_\mathcal{C} d\boldsymbol{\ell}\cdot\nabla \pot(\br)\, F(\br)
  = 0 ,
\end{equation}
where the last equality follows because
$d\boldsymbol{\ell}\cdot\nabla \pot(\br\in\mathcal{C})=0$, given that
the path $\mathcal{C}$ belongs completely to an
equipotential. Therefore, the field $n(\br)$ takes the same value on
all the points of this surface, which is thus also an isopycnic. As a
consequence, the value of the density on each isopycnic is determined
completely by the value of the potential on the associated
equipotential, and it is concluded that there must exist a functional
relationship, $n=\nu(\pot)$. A similar argument leads to the result
that a relationship $T=\tau(\pot)$ holds too.

\section{Weak--field limit around a homogeneous, stationary, quiescent state}
\label{sec:weakW}

Let us assume that, in the absence of an external field, the
hydrodynamic equations possess a unique stationary state that is
homogeneous and quiescent. Then, $\bu=0$ and the state is fully
characterized by a density $n^{(0)}$ and a temperature $T^{(0)}$, both
uniform in time and space and which must be related by the condition
for energy balance, see \eqs{eq:tempfull}:
\begin{equation}
  \label{eq:homstat}
  G^{(0)} := G(n^{(0)}, T^{(0)} ) = T^{(0)} \zeta(n^{(0)}, T^{(0)}) -  Q(n^{(0)}, T^{(0)}) = 0 .
\end{equation}
We assume that this equation provides a unique relationship, so that
the specification of $n^{(0)}$ gives a single stationary, homogeneous
and quiescent state. We further assume that this state is linearly
stable, which imposes constraints on the values of the parameters
\cite{BRS13}; see also after \eq{eq:M} below.  We introduce a
bookkeeping parameter $\varepsilon$ by the replacement
$\pot(\br) \rightarrow \varepsilon \pot(\br)$, that will be set equal
to unity at the end of the calculations. It allows one to seek a
solution of the hydrodynamic equations as an expansion in
$\varepsilon$ around the homogeneous and quiescent stationary state:
\begin{subequations}
  \label{eq:expansion}
    \begin{equation}
      n(\br, t) = n^{(0)} + \varepsilon n^{(1)} (\br, t) + \varepsilon^2 n^{(2)}
      (\br, t) + O(\varepsilon^3) ,
  \end{equation}
    \begin{equation}
      T(\br, t) = T^{(0)} + \varepsilon T^{(1)} (\br, t) + \varepsilon^2 T^{(2)}
      (\br, t) + O(\varepsilon^3) ,
  \end{equation}
    \begin{equation}
      \bu(\br, t) = \varepsilon \bu^{(1)} (\br, t) + \varepsilon^2 \bu^{(2)}
      (\br, t) + O(\varepsilon^3) .
    \end{equation}
\end{subequations}
Consistently with these expansions, we introduce also the following
auxiliary expressions (the subindices $n$, $T$ denote a derivative
with respect to $n$ or $T$, respectively, evaluated at the homogenous,
stationary state $n=n^{(0)}$, $T=T^{(0)}$):
\begin{subequations}
  \label{eq:coeffexp}
  \begin{equation}
    \label{eq:etaexp}
    \eta(n,T) = \eta^{(0)} + \varepsilon \eta^{(1)} + O(\varepsilon^2) 
    \qquad\Rightarrow\qquad
    \eta^{(1)} := \eta_n n^{(1)} + \eta_T T^{(1)} ,
  \end{equation}
  \begin{equation}
    \label{eq:kappaexp}
    \kappa(n,T) = \kappa^{(0)} + \varepsilon \kappa^{(1)} + O(\varepsilon^2) 
    \qquad\Rightarrow\qquad
    \kappa^{(1)} := \kappa_n n^{(1)} + \kappa_T T^{(1)} ,
  \end{equation}
  \begin{equation}
    \label{eq:muexp}
    \mu(n,T) = \mu^{(0)} + \varepsilon \mu^{(1)} + O(\varepsilon^2) 
    \qquad\Rightarrow\qquad
    \mu^{(1)} := \mu_n n^{(1)} + \mu_T T^{(1)} ,
  \end{equation}
  \begin{equation}
    \label{eq:pexp}
    p(n,T) = p^{(0)} + \varepsilon p^{(1)} + \varepsilon^2 p^{(2)} + O(\varepsilon^3) 
    \qquad\Rightarrow\qquad
    \left\{
      \begin{array}[c]{l}
        p^{(1)} := p_n n^{(1)} + p_T T^{(1)} , \\
        \\
        \begin{array}[t]{rcl}
          p^{(2)}
          & :=
          & \displaystyle p_n n^{(2)} + p_T T^{(2)} + \frac{1}{2} p_{nn} \left[n^{(1)}\right]^2
          \\
          \\
          &
          & \displaystyle \mbox{} + \frac{1}{2} p_{TT} \left[T^{(1)}\right]^2 + p_{nT} n^{(1)} T^{(1)} ,
        \end{array}
      \end{array}
    \right.
  \end{equation}
  \begin{equation}
    \label{eq:Gexp}
    G(n,T) = \varepsilon G^{(1)} + \varepsilon^2 G^{(2)} + O(\varepsilon^3) 
    \qquad\Rightarrow\qquad
    \left\{
      \begin{array}[c]{l}
        G^{(1)} := G_n n^{(1)} + G_T T^{(1)} , \\
        \\
        \begin{array}[t]{rcl}
          G^{(2)}
          & :=
          & \displaystyle G_n n^{(2)} + G_T T^{(2)} + \frac{1}{2}
            G_{nn} \left[n^{(1)}\right]^2
          \\
          \\
          &
          & \displaystyle \mbox{} + \frac{1}{2} G_{TT} \left[T^{(1)}\right]^2 + G_{nT} n^{(1)} T^{(1)} ,
        \end{array}
      \end{array}
    \right.
  \end{equation}
  \begin{equation}
    \label{eq:stressexp}
    \mathsf{\Sigma} = \varepsilon \mathsf{\Sigma}^{(1)}
    + \varepsilon^2 \mathsf{\Sigma}^{(2)} + O(\varepsilon^3)
    \qquad\Rightarrow\qquad
    \left\{
      \begin{array}[c]{lcl}
        \mathsf{\Sigma}^{(1)}
        & :=
        & \displaystyle \eta^{(0)} \left[ \nabla \bu^{(1)}
        + (\nabla \bu^{(1)})^\dagger 
          - \frac{2}{D} \mathsf{I}\, \nabla\cdot\bu^{(1)} \right] ,
        \\
        \\
        \mathsf{\Sigma}^{(2)}
        & :=
        & \displaystyle \eta^{(1)} \left[ \nabla \bu^{(1)}
          + (\nabla \bu^{(1)})^\dagger
          - \frac{2}{D} \mathsf{I}\, \nabla\cdot\bu^{(1)} \right]
        \\
        \\
        & 
        & \displaystyle + \eta^{(0)} \left[ \nabla \bu^{(2)}
          + (\nabla \bu^{(2)})^\dagger 
          - \frac{2}{D} \mathsf{I}\, \nabla\cdot\bu^{(2)} \right] ,
      \end{array}
    \right.
  \end{equation}
  \begin{equation}
    \label{eq:heatexp}
    \bq = \varepsilon \bq^{(1)}
    + \varepsilon^2 \bq^{(2)} + O(\varepsilon^3)
    \qquad\Rightarrow\qquad
    \left\{
      \begin{array}[c]{lcl}
        \bq^{(1)}
        & :=
        & \displaystyle -\kappa^{(0)} \nabla T^{(1)} - \mu^{(0)} \nabla n^{(1)} ,
        \\
        \\
        \bq^{(2)}
        & :=
        & \displaystyle -\kappa^{(1)} \nabla T^{(1)} - \mu^{(1)} \nabla n^{(1)}
          - \kappa^{(0)} \nabla T^{(2)} - \mu^{(0)} \nabla n^{(2)} .
      \end{array}
    \right.
  \end{equation}
\end{subequations}
Inserting these expansions in Eqs.~(\ref{eq:cont}--\ref{eq:temp}), one
gets
\begin{subequations}
  \label{eq:hydroexp}
  \begin{equation}
    \label{eq:contexp}
    \varepsilon \frac{\partial n^{(1)}}{\partial t}
    + \varepsilon^2 \frac{\partial n^{(2)}}{\partial t}
    = - \varepsilon n^{(0)} \nabla\cdot \bu^{(1)} - \varepsilon^2
    \left[ \nabla\cdot (n^{(1)} \bu^{(1)}) + n^{(0)} \nabla\cdot \bu^{(2)} \right]
    + O(\varepsilon^3) ,
  \end{equation}
  \begin{equation}
    \label{eq:velexp}
    \varepsilon m n^{(0)} \frac{\partial \bu^{(1)}}{\partial t}
    + \varepsilon^2 \left[
      m n^{(1)} \frac{\partial \bu^{(1)}}{\partial t} 
      + m n^{(0)} \frac{\partial \bu^{(2)}}{\partial t}
    \right]
    =
    \begin{array}[t]{cl}
      \displaystyle
      - & \varepsilon \left[ \nabla p^{(1)}  + n^{(0)} \nabla W -
      \nabla\cdot\mathsf{\Sigma}^{(1)} \right]
      \\
      \\
      - & \displaystyle \varepsilon^2 \left[
      m n^{(0)} (\bu^{(1)}\cdot\nabla)\bu^{(1)}
      + \nabla p^{(2)} + n^{(1)} \nabla W
      - \nabla\cdot\mathsf{\Sigma}^{(2)}
      \right] 
      \\
      \\
      + & O(\varepsilon^3) ,
    \end{array}
  \end{equation}
  \begin{equation}
    \label{eq:tempexp}
    \varepsilon \frac{D}{2} n^{(0)} \frac{\partial T^{(1)}}{\partial t}
    + \varepsilon^2 \frac{D}{2} \left[
      n^{(1)} \frac{\partial T^{(1)}}{\partial t}
      + n^{(0)} \frac{\partial T^{(2)}}{\partial t}
    \right]
    =
    \begin{array}[t]{cl}
      - & \displaystyle \varepsilon \left[ \frac{D}{2} n^{(0)}\, G^{(1)}
          + p^{(0)} \nabla\cdot\bu^{(1)}
          + \nabla\cdot\bq^{(1)}
        \right]
      \\
      \\
      - & \displaystyle \varepsilon^2 \left[
          (\bu^{(1)} \cdot \nabla) T^{(1)}
          + \frac{D}{2} \left( n^{(1)}\, G^{(1)} + n^{(0)}\, G^{(2)} \right)
          \right. \\ &\qquad \left.+ p^{(1)} \nabla\cdot\bu^{(1)}
          + p^{(0)} \nabla\cdot\bu^{(2)}
          - \mathsf{\Sigma}^{(1)} : (\nabla\bu^{(1)})
          + \nabla\cdot\bq^{(2)}
          \right]
      \\
      \\
      + & O(\varepsilon^3) .
    \end{array}
  \end{equation}
\end{subequations}
Upon collecting terms of the same order in $\varepsilon$, one arrives
at the following sets of linear equations:
\begin{itemize}
\item \textbf{First order in $\varepsilon$:} 
  \begin{subequations}
    \label{eq:hydro1}
    \begin{equation}
      \label{eq:cont1}
      \frac{\partial n^{(1)}}{\partial t} = - n^{(0)} \nabla\cdot \bu^{(1)} ,
    \end{equation}
    \begin{equation}
      \label{eq:vel1}
      m n^{(0)} \frac{\partial \bu^{(1)}}{\partial t} = - \nabla p^{(1)} - n^{(0)}
      \nabla\pot + \nabla\cdot \mathsf{\Sigma}^{(1)} ,
    \end{equation}
    \begin{equation}
      \label{eq:temp1}
      \frac{D}{2} n^{(0)} \frac{\partial T^{(1)}}{\partial t} 
      = - \frac{D}{2} n^{(0)} G^{(1)} - p^{(0)} \nabla\cdot\bu^{(1)}
      - \nabla\cdot\bq^{(1)} .
    \end{equation}
  \end{subequations}
  These equations can be written in a compact notation as
  \begin{equation}
    \label{eq:eps1}
    \frac{\partial}{\partial t} \left(
      \begin{array}[c]{c}
        n^{(1)} \\
        \bu^{(1)} \\
        T^{(1)}
      \end{array}
    \right)
    = \mathcal{M} \left(
      \begin{array}[c]{c}
        n^{(1)} \\
        \bu^{(1)} \\
        T^{(1)}
      \end{array}
    \right)
    -  \left(
      \begin{array}[c]{c}
        0 \\
        \displaystyle \frac{1}{m} \nabla\pot \\
        0
      \end{array}
    \right) ,
  \end{equation}
  in terms of the operator--valued hydrodynamic matrix
  \begin{equation}
    \label{eq:M}
    \mathcal{M} := \left(
      \begin{array}[c]{ccc}
        0 
        & - n^{(0)} \nabla\cdot 
        & 0 
        \\
        & & \\
        \displaystyle- \frac{p_n}{m n^{(0)}} \nabla 
        & \displaystyle \frac{\eta^{(0)}}{m n^{(0)}} \left[ \nabla^2 + \left(1-\frac{2}{D}
          \right) \nabla\nabla\cdot \right] 
        & \displaystyle - \frac{p_T}{m n^{(0)}} \nabla 
        \\
        & & \\
        \displaystyle - G_n + \frac{2}{D n^{(0)}} \mu^{(0)} \nabla^2
        & \displaystyle - \frac{2 p^{(0)}}{D n^{(0)}} \nabla\cdot
        & \displaystyle - G_T + \frac{2}{D n^{(0)}} \kappa^{(0)} \nabla^2
      \end{array}
    \right) ,
  \end{equation}
  as obtained after inserting the expansions~(\ref{eq:coeffexp}).  The
  linear stability of the homogeneous, stationary state is determined
  by the eigenvalues of the matrix $\mathcal{M}$: the state is stable
  if they all have a negative real part, which imposes restrictions on
  the parameters: see the detailed discussion in
  \rcite{BRS13}. Assuming this to be the case, and since $\pot(\br)$
  does not depend on time, the long--time solution of \eq{eq:eps1}
  will approach a stationary state given as the formal solution
  \begin{equation}
    \left(
      \begin{array}[c]{c}
        n^{(1)} \\
        \bu^{(1)} \\
        T^{(1)}
      \end{array}
    \right) _\mathrm{stat}
    = \mathcal{M}^{-1} \left(
      \begin{array}[c]{c}
        0 \\
        \displaystyle \frac{1}{m} \nabla\pot \\
        0
      \end{array}
    \right) .
  \end{equation}
  Notice that the stability assumption rules out the existence of a
  null eigenvalue, and thus $\mathcal{M}$ is indeed invertible (this
  reflects the physical assumption concerning \eq{eq:homstat} that, in
  the absence of an external potential ($\pot \equiv 0$), the
  stationary state $n=n^{(0)}, \bu=0, T=T^{(0)}$, that is,
  $n^{(1)}=\bu^{(1)}=T^{(1)}=0$ is unique).
  Equivalently, one can write for the stationary state directly from
  \eqs{eq:hydro1}, upon inserting Eqs.~(\ref{eq:stressexp},
  \ref{eq:heatexp}), 
  \begin{subequations}
    \begin{equation}
      \label{eq:incomp1}
      \nabla\cdot\bu^{(1)} = 0 ,
    \end{equation}
    \begin{equation}
      \label{eq:stokes1}
      \eta^{(0)} \nabla^2 \bu^{(1)} - \nabla p^{(1)} - n^{(0)} \nabla \pot = 0 ,
    \end{equation}
    \begin{equation}
      \kappa^{(0)} \nabla^2 T^{(1)} + \mu^{(0)} \nabla^2 n^{(1)} - \frac{D}{2}
        n^{(0)} G^{(1)} = 0 .
    \end{equation}
  \end{subequations}
  Taking the curl of \eq{eq:stokes1}, one gets
  $\nabla^2 (\nabla\times \bu^{(1)})=0$. Assuming neutral boundary
  conditions that do not excite flow, this last equation and
  \eq{eq:incomp1}, lead to
  \begin{subequations}
    \begin{equation}
      \label{eq:noflow1}
      \bu^{(1)} = 0 ,
    \end{equation}
    \begin{equation}
      \label{eq:mech1}
      \nabla p^{(1)} + n^{(0)} \nabla \pot = 0 ,
    \end{equation}
    \begin{equation}
      \label{eq:heat1}
      \kappa^{(0)} \nabla^2 T^{(1)} + \mu^{(0)} \nabla^2 n^{(1)} - \frac{D}{2}
        n^{(0)} G^{(1)} = 0 ,
    \end{equation}
  \end{subequations}
  from where the stationary density and temperature profiles can be
  obtained to first order in $\pot$.

\item \textbf{Second order in $\varepsilon$:} upon making use of
  \eqs{eq:hydro1} in order to eliminate $\partial \bu^{(1)}/\partial t$,
  $\partial T^{(1)}/\partial t$, one gets
  \begin{subequations}
    \label{eq:hydro2}
    \begin{equation}
      \label{eq:cont2}
      \frac{\partial n^{(2)}}{\partial t} = - n^{(0)} \nabla\cdot \bu^{(2)} -
      \nabla\cdot (n^{(1)} \bu^{(1)}) ,
    \end{equation}
    \begin{eqnarray}
      \label{eq:vel2}
      m n^{(0)} \frac{\partial \bu^{(2)}}{\partial t}
      = - m n^{(0)} (\bu^{(1)}\cdot\nabla)\bu^{(1)} - \nabla p^{(2)}
        + \frac{n^{(1)}}{n^{(0)}} \nabla p^{(1)}
        + \nabla\cdot \mathsf{\Sigma}^{(2)}
        - \frac{n^{(1)}}{n^{(0)}} \nabla\cdot \mathsf{\Sigma}^{(1)} ,        
    \end{eqnarray}
    \begin{eqnarray}
      \label{eq:temp2}
      \frac{D}{2} n^{(0)} \frac{\partial T^{(2)}}{\partial t} 
      & = 
      & - \frac{D}{2} n^{(0)} G^{(2)} - p^{(0)} \nabla\cdot\bu^{(2)}
        - \nabla\cdot\bq^{(2)}
        \nonumber
      \\
      \nonumber \\
      & 
      & - (\bu^{(1)}\cdot\nabla) T^{(1)} 
        - p^{(1)} \nabla\cdot\bu^{(1)}
        + \mathsf{\Sigma}^{(1)} : (\nabla\bu^{(1)}) 
        \nonumber
      \\
      \nonumber \\
      & 
      & + \frac{n^{(1)}}{n^{(0)}} \left[
        p^{(0)} \nabla\cdot\bu^{(1)} 
        + \nabla\bq^{(1)}
        \right] .
    \end{eqnarray}
  \end{subequations}
  These equations can be written also in compact form by means of the
  hydrodynamic matrix~(\ref{eq:M}):
  \begin{equation}
    \label{eq:eps2}
    \frac{\partial}{\partial t} \left(
      \begin{array}[c]{c}
        n^{(2)} \\
        \bu^{(2)} \\
        T^{(2)}
      \end{array}
    \right)
    = \mathcal{M} \left(
      \begin{array}[c]{c}
        n^{(2)} \\
        \bu^{(2)} \\
        T^{(2)}
      \end{array}
    \right)
    + \left(
      \begin{array}[c]{c}
        S_\mathrm{dens} \\
        S_\mathrm{vel} \\
        S_\mathrm{temp}
      \end{array}
      \right) ,
  \end{equation}
  with certain source terms
  $S_\mathrm{dens},S_\mathrm{vel}, S_\mathrm{temp}$ that are quadratic
  in the first-order fields $n^{(1)}, \bu^{(1)}, T^{(1)}$, and we do
  not need to write down explicitly. The long--time solution of
  \eq{eq:eps2} will again approach a stationary state, given in terms
  of the long--time sources formally as
  \begin{equation}
        \left(
      \begin{array}[c]{c}
        n^{(2)} \\
        \bu^{(2)} \\
        T^{(2)}
      \end{array}
    \right)_\mathrm{stat}
    = - \mathcal{M}^{-1} \left(
      \begin{array}[c]{c}
        S_\mathrm{dens}  (t\to+\infty) \\
        S_\mathrm{vel}  (t\to+\infty) \\
        S_\mathrm{temp}  (t\to+\infty) 
      \end{array}
    \right) .
  \end{equation}
  The fields $n^{(2)}$, $T^{(2)}$ would introduce a quantitative
  correction to the first order perturbations $n^{(1)}$,
  $T^{(1)}$. The most interesting result here is actually the
  non-vanishing velocity field $\bu^{(2)}$, which represents a
  qualitative difference with both the first--order corrections and
  the equilibrium state of an elastic fluid. Thus, we focus on this
  velocity field and, after simplifying for a stationary state, we get
  the Stokes equations for incompressible flow excited by the external
  potential $\pot$:
  \begin{subequations}
    \label{eq:flow2}
    \begin{eqnarray}
      \label{eq:divu2}
      \textrm{\eq{eq:cont2}}
      & \qquad\stackrel{\textrm{\tiny (\ref{eq:noflow1})}}{\Rightarrow}\qquad 
      & \nabla\cdot\bu^{(2)} = 0 ,
      \\
      \label{eq:stokesu2}
        \textrm{\eq{eq:vel2}}
      & \qquad\stackrel{\textrm{\tiny (\ref{eq:mech1}, \ref{eq:stressexp})}}{\Rightarrow}\qquad 
      & \nabla\cdot \mathsf{\Sigma}^{(2)}
        = \eta^{(0)} \nabla^2 \bu^{(2)}
        = \nabla p^{(2)} +
        n^{(1)} \nabla \pot .
    \end{eqnarray}
  \end{subequations}

\end{itemize}

\subsection{The perturbative equations~(\ref{eq:plin}--\ref{eq:flowlin})}
\label{sec:pert}

The relevant expressions are Eqs.~(\ref{eq:mech1}, \ref{eq:heat1}) for
the density and temperature profiles, and \eqs{eq:flow2} for the
velocity field. When the first one is expanded with the help of
\eq{eq:pexp}, one arrives at \eq{eq:plin} in the manuscript,
\begin{equation}
  \label{eq:mech1bis}
  \nabla p^{(1)} = p_n \nabla n^{(1)}  + p_T \nabla T^{(1)} = - n^{(0)} \nabla \pot .
\end{equation}
This equation can be integrated,
\begin{equation}
  \label{eq:hydrostatic1} 
  p_n \, n^{(1)}(\br) + p_T \, T^{(1)}(\br) = - n^{(0)} \pot(\br) ,
\end{equation}
where an integration constant has been absorbed in the potential
$\pot$. This expression allows one to eliminate one of the fields
($n^{(1)}$ or $T^{(1)}$) in favor of the other in \eq{eq:heat1}; for
instance, when the definition~(\ref{eq:Gexp}) is employed:
\begin{equation}
  \left( p_n \kappa^{(0)} - p_T \mu^{(0)}\right) \nabla^2 n^{(1)}
  - \frac{D}{2} n^{(0)} \left( p_n G_T - p_T G_n\right) n^{(1)}
  = - n^{(0)} \kappa^{(0)} \nabla^2 \pot 
  + \frac{D}{2} \left[n^{(0)}\right]^2 G_T \pot . 
\end{equation}
And this equation is then transformed into \eq{eq:heatlin} of the
manuscript,
\begin{equation}
  \corr^2 \nabla^2 \rho - \rho 
  = \gamma \pot ,
\end{equation}
in terms of the auxiliary field
\begin{equation}
  \label{eq:rhodef}
  \rho(\br) :=
  n^{(1)}(\br) + \frac{n^{(0)}\kappa^{(0)}}{p_n \kappa^{(0)} - p_T
    \mu^{(0)}} \pot(\br)
  \stackrel{\textrm{\tiny (\ref{eq:hydrostatic1})}}{=} \frac{p_T}{p_n\kappa^{(0)} - p_T\mu^{(0)}} \left[
    - \mu^{(0)} n^{(1)}(\br) - \kappa^{(0)} T^{(1)}(\br)
  \right] ,
\end{equation}
see \eq{eq:rho}, and the auxiliary quantities
\begin{equation}
  \label{eq:corrSM}
  \corr^2 := \frac{2}{D n^{(0)}} \, \frac{p_n \kappa^{(0)} - p_T \mu^{(0)}}{p_n G_T
    - p_T G_n},
\end{equation}
\begin{equation}
  \label{eq:gammaSM}
  \gamma
  := \frac{D n^{(0)} \corr^2}{2 (p_n \kappa^{(0)} - p_T \mu^{(0)})} \left[
    n^{(0)} G_T
    - \frac{n^{(0)} \kappa^{(0)} (p_n G_T- p_T G_n)}{p_n \kappa^{(0)} - p_T \mu^{(0)}}
  \right]
  = \frac{n^{(0)} p_T}{p_n \kappa^{(0)} - p_T \mu^{(0)}}\;
  \frac{\kappa^{(0)} G_n - \mu^{(0)} G_T}{p_n G_T
    - p_T G_n} ,
\end{equation}
see Eqs.~(\ref{eq:corr}, \ref{eq:gamma}).  The field $\rho$ is closely
related to the heat current to first order, see \eq{eq:heatexp}:
\begin{equation}
  \label{eq:rhoq}
  \nabla\rho = \frac{p_T}{p_n \kappa^{(0)} - p_T \mu^{(0)}} \bq^{(1)} .
\end{equation}
(As we shall see in the following, the absolute value of $\rho$, and
thus the value of an additive constant in $\pot$, is actually
irrelevant for the physical predictions, as they will eventually
depend only on $\nabla \rho$.) The length scale $\corr$ is real
because the homogeneous and quiescent stationary state is assumed
stable, which implies $G_T>0$, $p_n G_T - p_T G_n>0$ \cite{BRS13}, and
because we can restrict the model to sufficiently small inelasticities
that $p_n \kappa^{(0)} - p_T \mu^{(0)}>0$, so that heat flows against
temperature gradients at zero potential:
\begin{equation}
  -\nabla (\kappa^{(0)} T^{(1)} + \mu^{(0)} n^{(1)})
  \;\stackrel{\textrm{\tiny (\ref{eq:mech1bis})}}{=}\;
  \left( - \kappa^{(0)} + \frac{p_T}{p_n} \mu^{(0)} \right) \nabla T^{(1)}
  + \frac{n^{(0)} \mu^{(0)}}{p_n} \nabla\pot 
  \quad\stackrel{W\to 0}{\longrightarrow}\quad
  - \frac{1}{p_n} \left( p_n \kappa^{(0)} - p_T \mu^{(0)} \right) \nabla T^{(1)} .
\end{equation}
Finally, in \eq{eq:stokesu2} the second--order pressure term $p^{(2)}$
plays only a subordinate role as enforcer of the incompressibility
constraint~(\ref{eq:divu2}); it can be eliminated by taking the curl
of \eq{eq:stokesu2}, and this leads to Eqs.~(\ref{eq:divu},
\ref{eq:flowlin}) of the manuscript respectively,
\begin{equation}
  \nabla\cdot\bu^{(2)} = 0 ,
\end{equation}
\begin{equation}
  \eta^{(0)} \nabla^2 (\nabla\times\bu^{(2)})
  = \nabla n^{(1)} \times \nabla\pot
  \;\stackrel{\textrm{\tiny (\ref{eq:mech1bis})}}{=}\;
  - \frac{1}{n^{(0)}} \nabla n^{(1)} \times \nabla p^{(1)}
  \;\stackrel{\textrm{\tiny (\ref{eq:rhodef})}}{=}\;
  \nabla \rho \times \nabla\pot .
\end{equation}
That is, only the solenoidal component of $n^{(1)}\nabla\pot$ in
\eq{eq:stokesu2} drives the flow, and any potential component is
balanced by the pressure $p^{(2)}$.  This way of writing the equations
emphasizes the relevance of baroclinity.

\section{Solving the perturbative equations in a bounded domain}
\label{sec:bounded}

Here we solve Eqs.~(\ref{eq:heatlin}--\ref{eq:flowlin}) in a finite
two--dimensional ($D=2$) domain with periodic boundary conditions. We
consider a rectangular box (sidelengths $L_x, L_y$) in the
$XY$--plane. In this case, the flow reduces to
\begin{equation}
  \label{eq:2}
  \bu(\br) = u_x(x,y) \be_x + u_y(x,y) \be_y , 
\end{equation}
and it can be represented by means of a stream function
$S(x,y)$ as follows:
\begin{equation}
  \label{eq:stream}
  \nabla S := \bu(x,y) \times \be_z
  \quad\Leftrightarrow\quad
  \bu(x,y) = \be_z\times\nabla S .
\end{equation}
The incompressibility constraint~(\ref{eq:divu}) is then satisfied
automatically. The stream function verifies a Poisson equation
that relates it to the vorticity as
\begin{equation}
  \label{eq:vorticity}
  \omega(x,y) := \be_z \cdot (\nabla\times\bu)
  = \nabla^2 S ,
\end{equation}
while the vorticity satisfies another Poisson
equation, see \eq{eq:flowlin}:
\begin{equation}
  \label{eq:vortexPoisson}
  \nabla^2 \omega = B(x,y), 
\end{equation}
where the baroclinic source of vorticity is
\begin{equation}
  \label{eq:Bsource}
  B(x,y) := \frac{1}{\eta^{(0)}}
  \be_z\cdot\left(\nabla\rho\times\nabla W\right) .
\end{equation}
\begin{figure}[t]
  \hfill\begin{tabular}[c]{c}
    $\pot/\pot_0$ 
    \\
    \includegraphics[width=.05\textwidth]{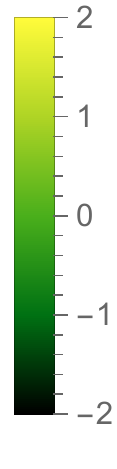}
  \end{tabular}
  \begin{tabular}[c]{c}
    \includegraphics[width=.4\textwidth]{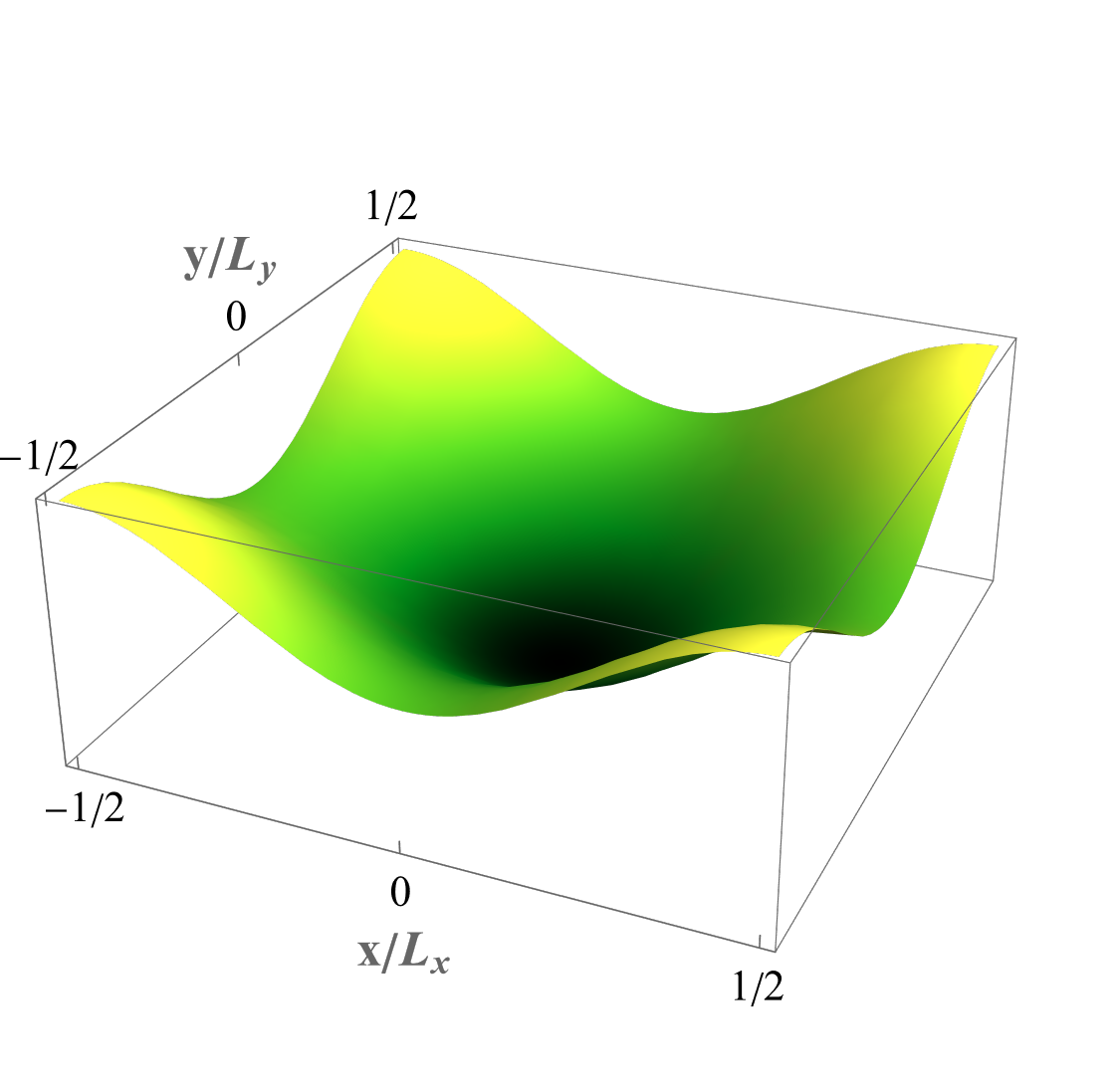}
  \end{tabular}
  \hfill
  \begin{tabular}[c]{c}
    \includegraphics[width=.4\textwidth]{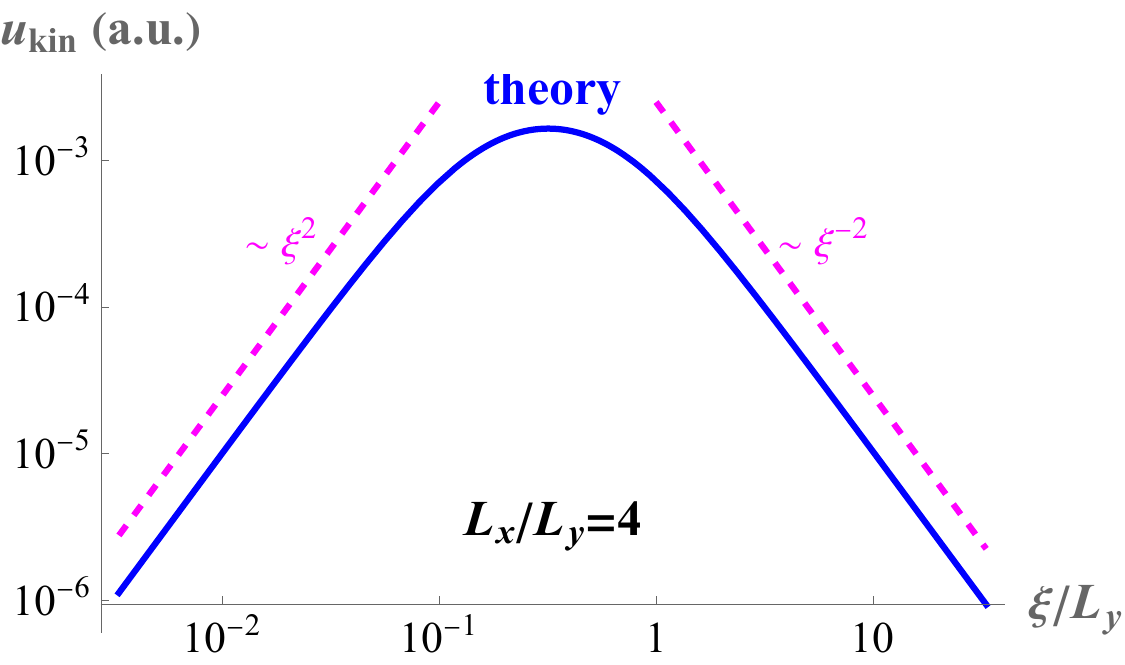}
  \end{tabular}
  \hspace*{\fill}
  \caption{(Left) The external potential~(\ref{eq:Wpart}). (Right) The
    plot of \eq{eq:ukinpart} (in arbitrary units) as a function of $\corr$, compared with
    the predicted asymptotic scalings.}
  \label{fig:Wpart}
\end{figure}
In view of the periodic boundary conditions, the fields can be
expanded in Fourier series: in terms of the orthonormal basis
\begin{equation}
  \label{eq:psi}
  \psi_{\alpha\beta}(x,y) := \frac{1}{\sqrt{L_x \, L_y }} \mathrm{e}^{2\pi i
    \left( \frac{\alpha x}{L_x} + \frac{\beta y}{L_y} \right) } ,
  \qquad 
  \alpha,\beta = 0, \pm 1, \pm 2, \dots
\end{equation}
one can write
\begin{equation}
  \label{eq:Wnm}
  \pot(x,y) = \sum_{\alpha,\beta=-\infty}^\infty\hat{\pot}_{\alpha\beta}
  \psi_{\alpha\beta}(x,y) ,
  \qquad
  \hat{\pot}_{\alpha\beta} = \int_{-L_x/2}^{L_x/2} dx\;
  \int_{-L_y/2}^{L_y/2} dy\;
  \psi_{\alpha\beta}^* (x,y)  \, \pot(x,y) ,
  \qquad
  \hat{W}_{\alpha\beta}^* = \hat{W}_{-\alpha,-\beta} , 
\end{equation}
and likewise for $\rho, \omega, S, B$. The solution to \eq{eq:heatlin}
is straightforward, since the Fourier basis consists of eigenfunctions
of the Laplacian with periodic boundary conditions:
\begin{equation}
  \label{eq:rhonm}
  \hat{\rho}_{\alpha\beta} = - \frac{\gamma \hat{W}_{\alpha\beta}}{1
    + \left(2\pi \alpha \corr/L_x\right)^2
    + \left(2\pi \beta \corr/L_y\right)^2
  } .
\end{equation}
One notices immediately that a potential field that would consist of a
single Fourier mode would not induce flow because then
$\rho \propto\pot$ and, consequently, $B=0$. In general, however, the
baroclinic source~(\ref{eq:Bsource}) is
\begin{eqnarray}
  \label{eq:Bnm}
  \hat{B}_{\alpha\beta}
  & =
  & - \frac{(2\pi)^2}{(L_x \, L_y)^{3/2} \, \eta^{(0)}}
  \sum_{\nu,\mu=-\infty}^\infty \left[ \nu (\alpha-\nu)- \mu (\beta-\mu) \right]
  \, \hat{\rho}_{\nu\mu} \, \hat{\pot}_{\alpha-\nu,\beta-\mu}
  \nonumber \\
  & =
  & \frac{(2\pi)^2 \, \gamma}{(L_x \, L_y)^{3/2} \, \eta^{(0)}}
  \sum_{\nu,\mu=-\infty}^\infty \frac{ \nu (\alpha-\nu) - \mu (\beta-\mu) }{1
    + \left(2\pi \nu \corr/L_x\right)^2
    + \left(2\pi \mu \corr/L_y\right)^2
  } \, \hat{\pot}_{\nu\mu} \, \hat{\pot}_{\alpha-\nu,\beta-\mu} .
\end{eqnarray}
The vorticity, the stream function and the flow field then follow from
Eqs.~(\ref{eq:stream}--\ref{eq:vortexPoisson}):
\begin{equation}
  \label{eq:omeganm}
  \hat{\omega}_{\alpha\beta} = - \frac{\hat{B}_{\alpha\beta}}{\left(2\pi \alpha/L_x\right)^2 +
    \left(2\pi \beta/L_y\right)^2},
\end{equation}
\begin{equation}
  \label{eq:Snm}
  \hat{S}_{\alpha\beta} = - \frac{\hat{\omega}_{\alpha\beta}}{\left(2\pi \alpha/L_x\right)^2 +
    \left(2\pi \beta/L_y\right)^2} , 
\end{equation}
\begin{equation}
  \label{eq:unm}
  \hat{\bu}_{\alpha\beta} = \left( -\be_x \frac{2\pi i \beta}{L_y} + \be_y
    \frac{2\pi i \alpha}{L_x} \right) \hat{S}_{\alpha\beta} .
\end{equation}
Finally, the total kinetic energy of the flow is
\begin{equation}
  \label{eq:totalK}
  K := \int_{-L_x/2}^{L_x/2} dx\; \int_{-L_y/2}^{L_y/2} dy \;
  \frac{1}{2} m \, n^{(0)} |\bu|^2
  = \frac{1}{2} m \, n^{(0)} \sum_{\alpha,\beta=-\infty}^\infty \left|
    \hat{\bu}_{\alpha\beta} \right|^2 ,
\end{equation}
from where a charateristic velocity can be defined as
\begin{equation}
  \label{eq:ukin}
  u_\mathrm{kin} := \sqrt{\frac{2 K}{m \, N}} ,
  \qquad
  N := n^{(0)} L_x\, L_y = \textrm{number of grains},
\end{equation}
which allows one to quantify the strength of the flow and to compare
the theoretical predictions with the simulations on a unified
framework.

\begin{figure}[t]
  \centering
  \hfill
  \begin{tabular}[c]{c}
    $|\bu|/u_\mathrm{kin}$
    \\
    \includegraphics[width=.05\textwidth]{figures/periodicBC/flow_colorBar.pdf}
  \end{tabular}
  \hfill
  \begin{tabular}[c]{cc}
    MD simulations, $L_x/L_y=1.6$, $N=40$
    & perturbative theory~(\ref{eq:upart}), $L_x/L_y=1.6$ 
    \\
    \includegraphics[width=.4\textwidth]{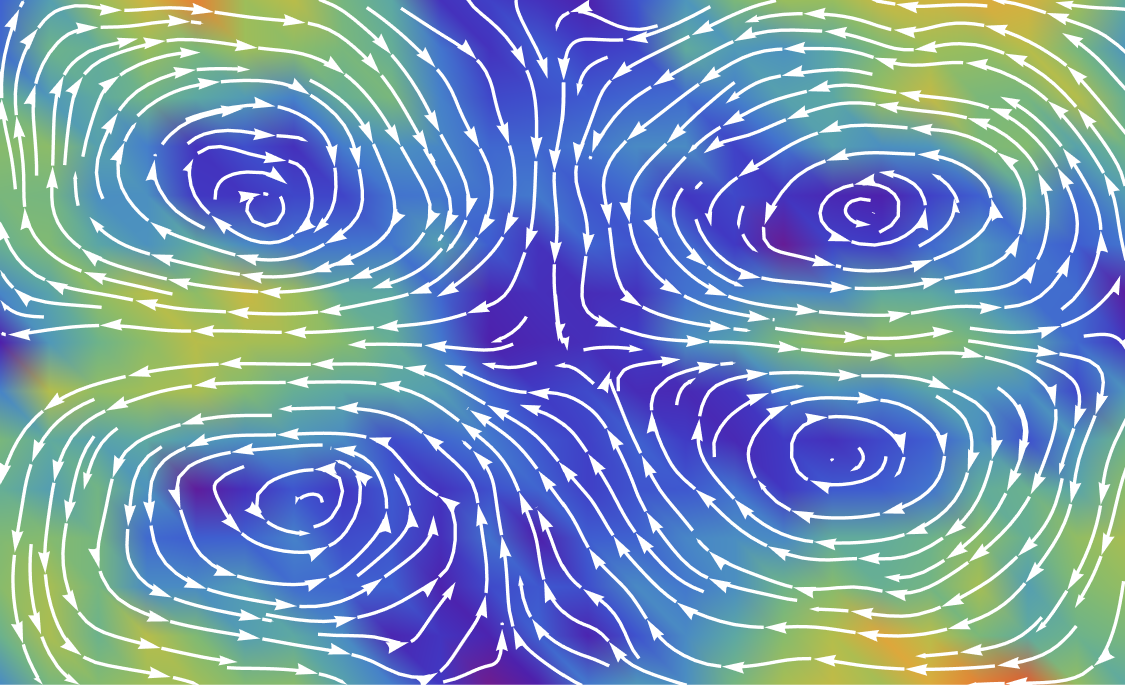}
    & \includegraphics[width=.4\textwidth]{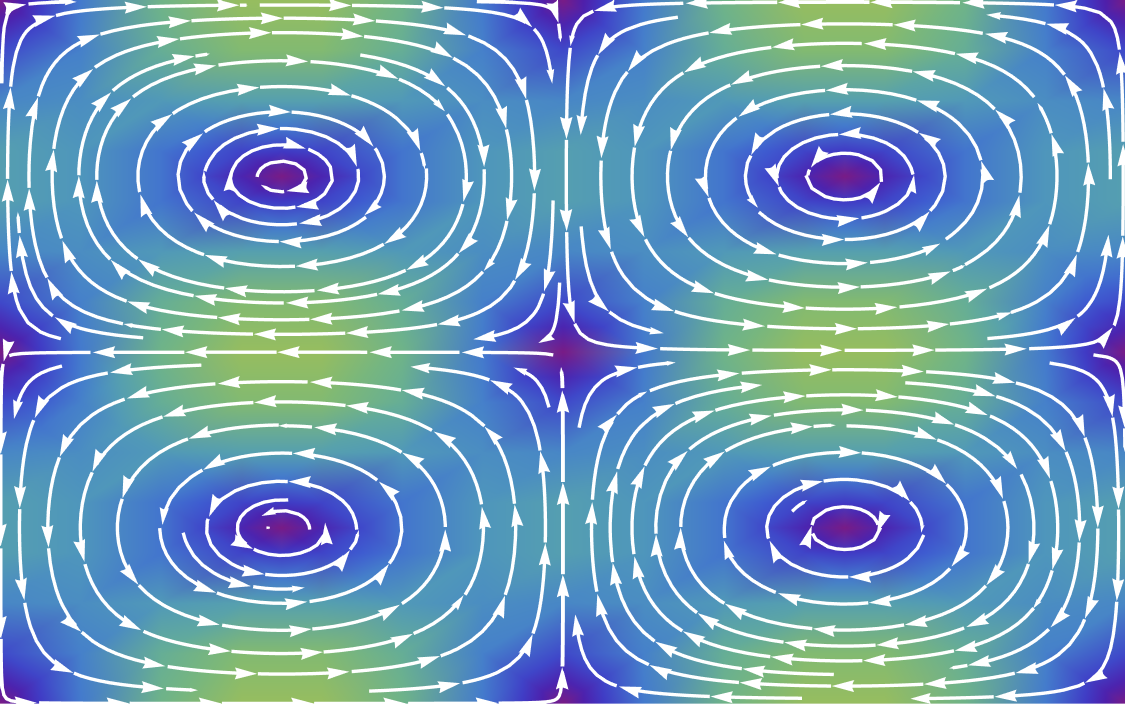}
    \\
    \\
    MD simulations, $L_x/L_y=2.0$, $N=50$
    & perturbative theory~(\ref{eq:upart}), $L_x/L_y=2.0$
    \\
    \includegraphics[width=.4\textwidth]{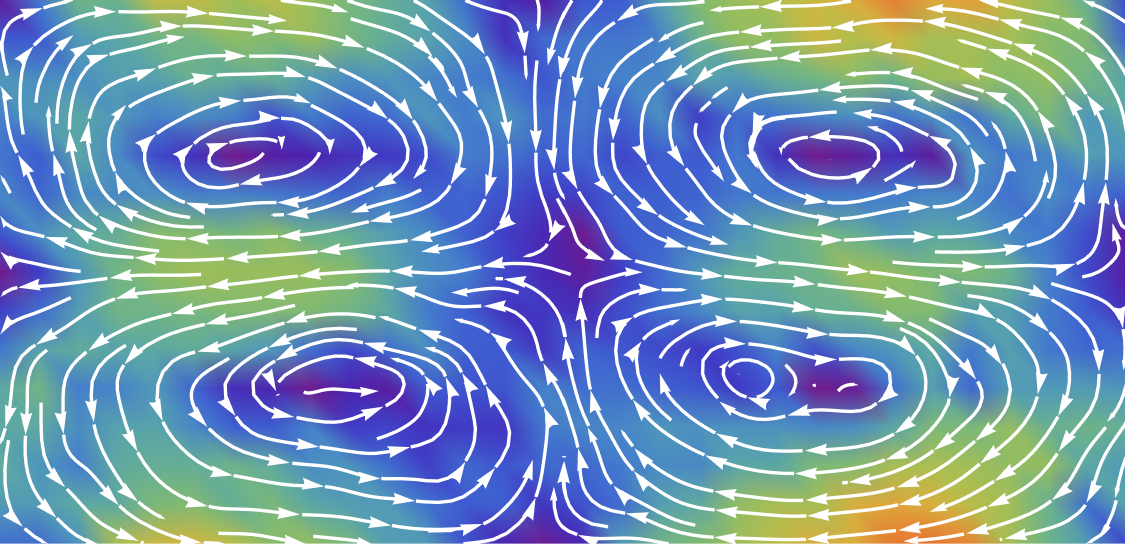}
    & \includegraphics[width=.4\textwidth]{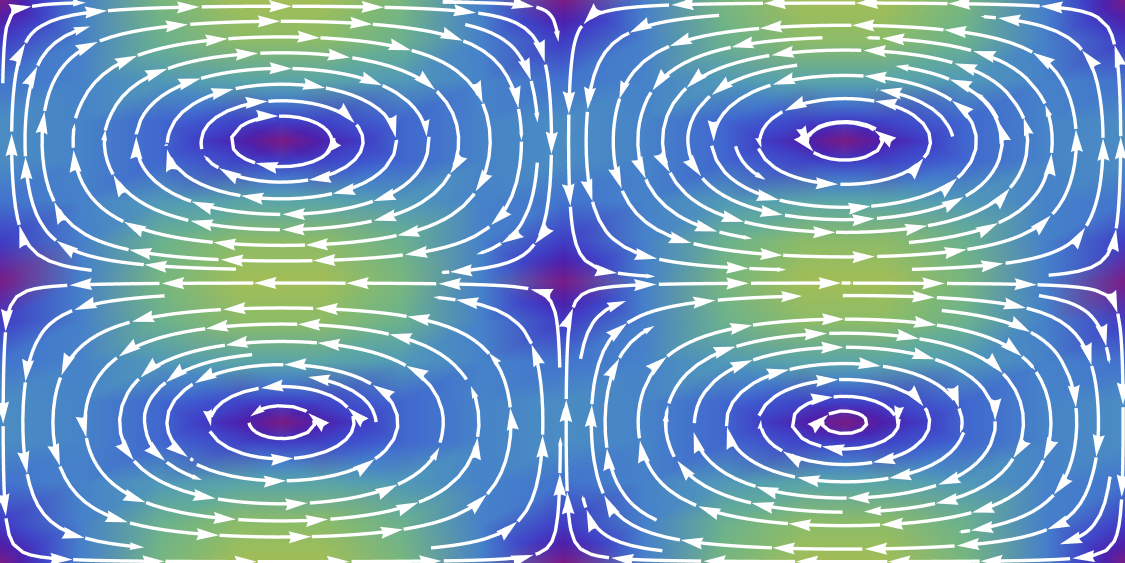}
    \\
    \\
    MD simulations, $L_x/L_y=3.0$, $N=75$
    & perturbative theory~(\ref{eq:upart}), $L_x/L_y=3.0$
    \\
    \includegraphics[width=.4\textwidth]{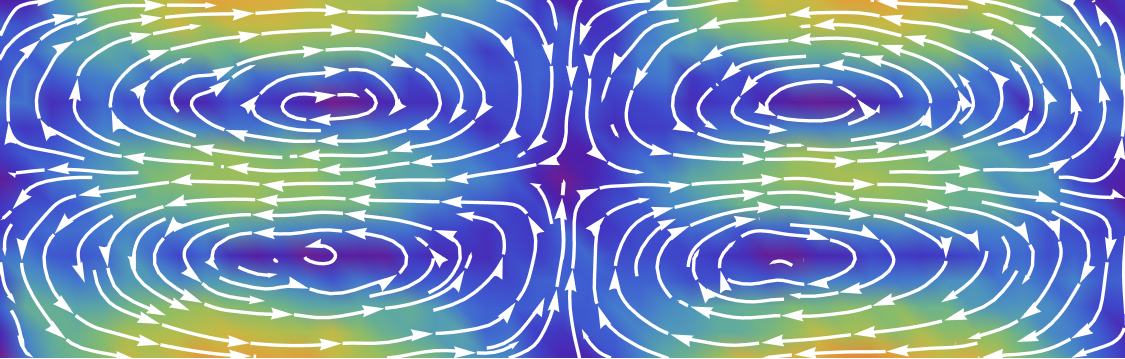}
    & \includegraphics[width=.4\textwidth]{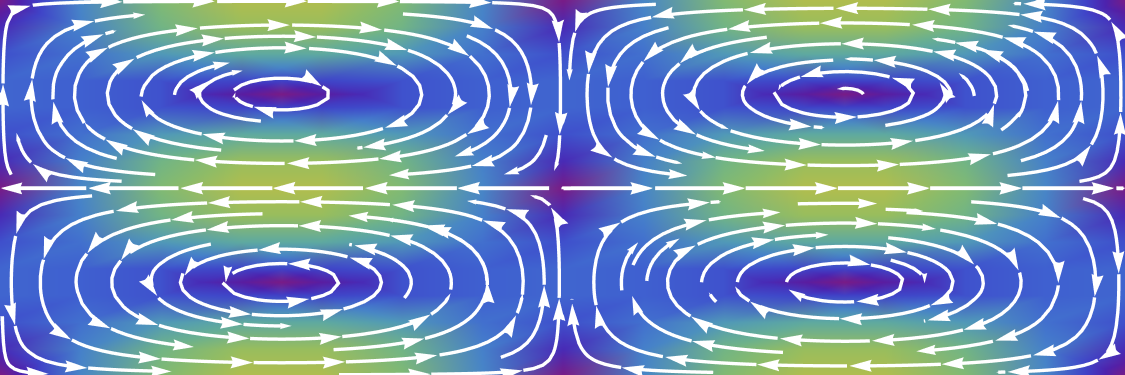}
    \\
    \\
    MD simulations, $L_x/L_y=4.0$, $N=100$
    & perturbative theory~(\ref{eq:upart}), $L_x/L_y=4.0$
    \\
    \includegraphics[width=.4\textwidth]{figures/periodicBC/flow_sim_Lx120.pdf}
    & \includegraphics[width=.4\textwidth]{figures/periodicBC/flow_theory_Lx120.pdf}
  \end{tabular}
  \hspace*{\fill}
  \caption{Comparison between the flows observed in Molecular Dynamics
    simulations (left column) and the perturbative prediction given by
    \eq{eq:upart} (right column).}
  \label{fig:flows}
\end{figure}

\begin{figure}[t]
  \centering
  \hfill
  \begin{tabular}[c]{c}
    $|\bu|/u_\mathrm{kin}$
    \\
    \includegraphics[width=.05\textwidth]{figures/periodicBC/flow_colorBar.pdf}
  \end{tabular}
  \begin{tabular}[c]{c}
    \includegraphics[width=.4\textwidth]{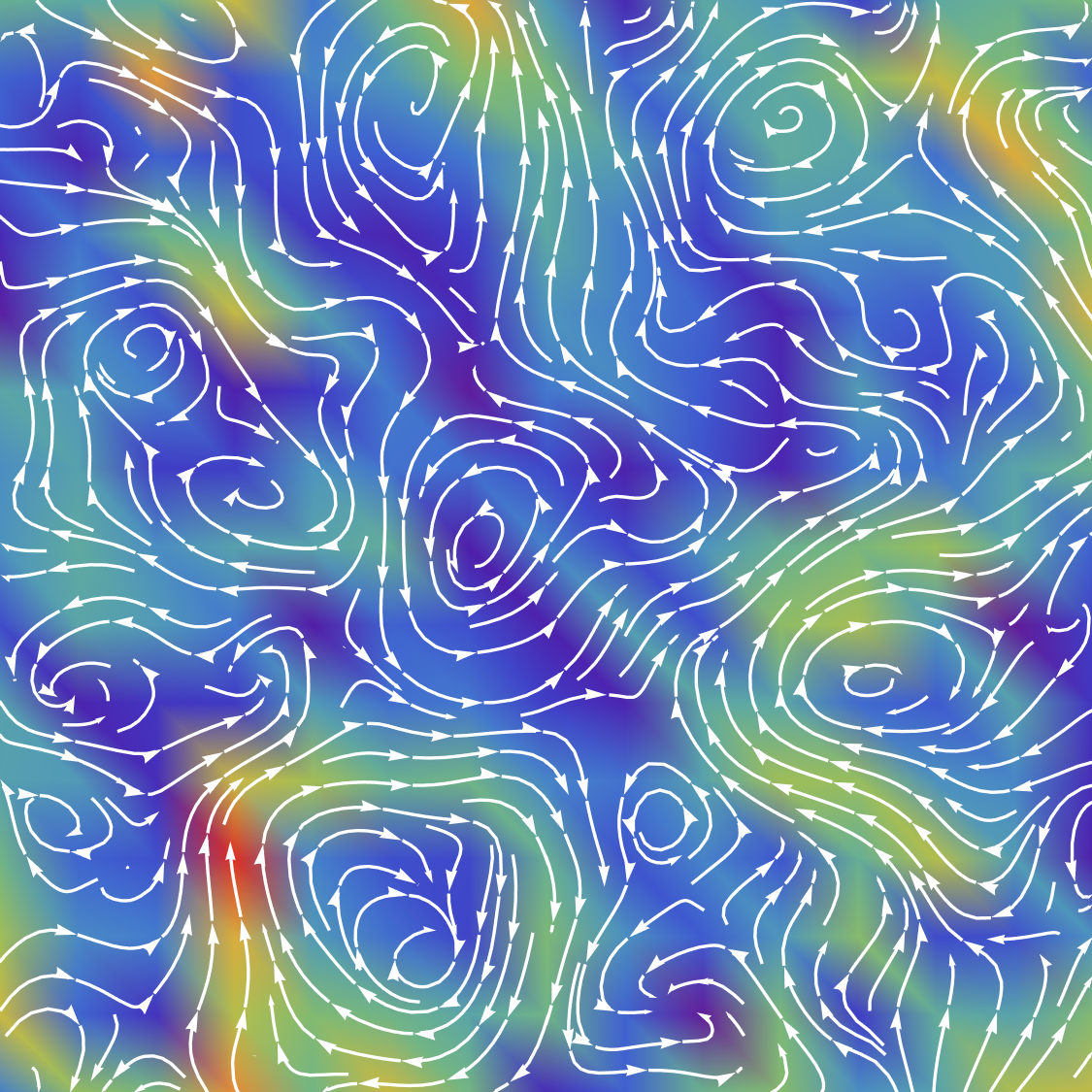}
  \end{tabular}
  \hspace*{\fill}
  \caption{The flow measured in simulations for the particular case of
    a square box ($L_x=L_y$).}
  \label{fig:flowSimSquareBox}
\end{figure}

These expressions are now particularized for the choice of the
external potential which consists of a superposition of the
fundamental modes in each direction (see \fref{fig:Wpart}, left;
$\pot_0$ is a given constant, which can be set equal to unity without
loss of generality):
\begin{equation}
  \label{eq:Wpart}
  \pot(x,y) =\mbox{} - \pot_0 \left( \cos \frac{2\pi x}{L_x} + \cos \frac{2\pi
      y}{L_y} \right)
  \quad\Rightarrow\quad
  \hat{\pot}_{\alpha\beta} = \frac{\pot_0 \sqrt{L_x\,L_y}}{2}
  \times \left\{
    \begin{array}[c]{cl}
      \mbox{} - 1 ,
      & (\alpha,\beta) = (\pm 1, 0) \;\mathrm{or}\; (0,\pm 1) ,
      \\
      \\
      0 ,
      & \mathrm{otherwise}.
      \\
    \end{array}
  \right.  
\end{equation}
This leads to a characteristic velocity scale 
\begin{equation}
  \label{eq:ukinpart}
  u_\mathrm{kin} = \frac{\pi |\gamma| \pot_0^2}{\eta^{(0)}}
  \frac{\corr^2 \, |L_x^2-L_y^2|}{\left( L_x^2+L_y^2 \right)^{3/2} \left[
      1 + \left(2\pi \corr/L_x\right)^2
    \right]
    \left[
      1 + \left(2\pi \corr/L_y\right)^2
    \right]
  } ,
\end{equation}
and to a flow of the form
\begin{equation}
  \label{eq:upart}
  \frac{\bu(x,y)}{u_\mathrm{kin}}
  = 2\; \mathrm{sign}(\gamma)\; \frac{\mathrm{sign}(L_x^2-L_y^2)}{\sqrt{L_x^2+L_y^2}}
  \left[
    \be_x L_x \sin\left(\frac{2\pi x}{L_x}\right) \, \cos\left(\frac{2\pi y}{L_y}\right)
    - \be_y L_y \cos\left(\frac{2\pi x}{L_x}\right) \, \sin\left(\frac{2\pi y}{L_y}\right)
  \right] .
\end{equation}
This result shows explicitly that the potential given by \eq{eq:Wpart}
is too symmetric to induce flow in a square box ($L_x=L_y$): in such
case, $\pot$ satisfies $\nabla^2 \pot \propto \pot$, which is the
linearized version of the generic constraint~(\ref{eq:stat3}). As the
aspect ratio $L_x/L_y$ changes from unity, the flow strength grows
because the asymmetry between the two fundamental modes becomes more
and more relevant, see \fref{fig:sim}. As a function of $\corr$ at
fixed $L_x, L_y$, one can also derive with ease the asymptotic
behaviors and the existence of a local maximum, see \fref{fig:Wpart},
in agreement with the generic argument provided in the main text.

This prediction can be compared with the flows measured in Molecular
Dynamics simulations (see \sref{sec:simul} for the details on the
implementation of the simulations). Figure~\ref{fig:flows} compares
these flows with the evaluation of the theoretical
expression~(\ref{eq:upart}) for different values of the aspect ratio
additional to those shown in \fref{fig:sim}. For completeness,
\fref{fig:flowSimSquareBox} shows the result from simulations for the
particular case of a square domain ($L_x=L_y$): one finds a noisy flow
pattern with no overtly recognizable symmetry and the smallest
recorded value of $u_\mathrm{kin}$.

\section{Self--phoresis}
\label{sec:self-phoresis}

\begin{figure}[t]
  \centering
  \hfill\includegraphics[width=.5\textwidth]{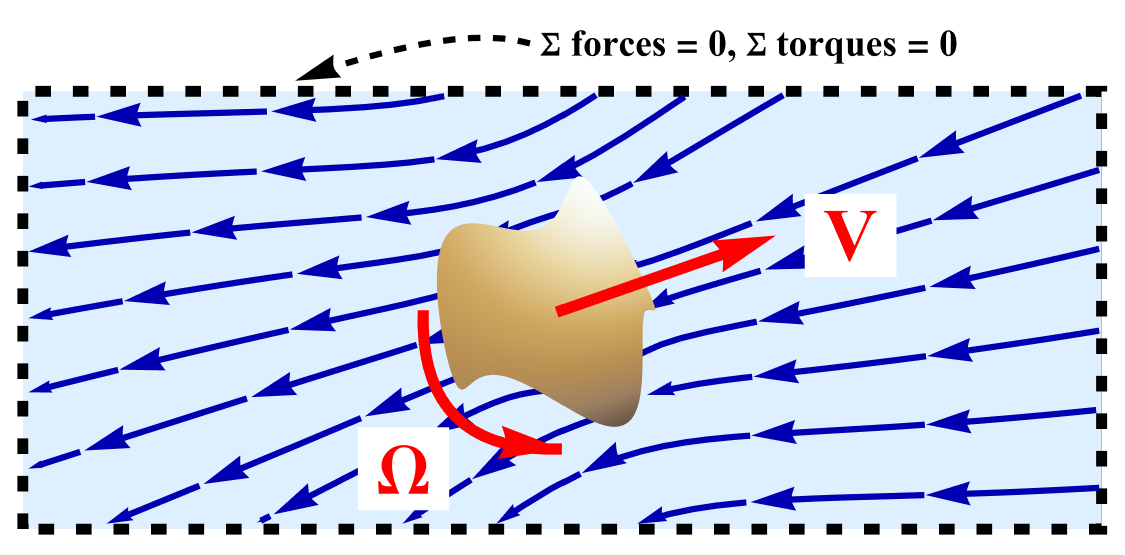}\hspace*{\fill}
  \caption{Sketch of phoresis. In spite of the system
    ``particle+fluid'' being mechanically isolated (depicted by the
    ``distant'' dashed border), there is directed motion (translation
    and rotation, red arrows) of an arbitrarily shaped particle
    (brown), and a corresponding flow (blue arrows) ensues in the
    fluid medium to conserve linear and angular momenta. The brownish
    nuances on the particle depict variations on physical and chemical
    properties of its surface, e.g., interaction potential with the
    fluid, surface charge, catalytic activity, etc.}
  \label{fig:phoresisSketch}
\end{figure}

Here we address the case that the potential $\pot(\br)$ is generated
exclusively by a free intruder in an unbounded three--dimensional
($D=3$) domain. It will be found that the intruder exhibits direct
motion (translation and rotation), formally analogous to self-phoresis
of a chemically active particle. For the latter, we follow
\rcites{DSZK47,Ande89,DPRD20,DoPo22,DoPo24a,DoPo24b} closely and
briefly review the modelling. Phoresis describes the directed (i.e.,
as opposed to Brownian) motion of a particle immersed in a fluid
\textit{when the system ``particle+fluid'' is mechanically isolated},
see \fref{fig:phoresisSketch}. This is an intrinsically
out-of-equilibrium phenomenon (a net particle current violates
detailed balance), and the feature of mechanical isolation makes the
problem fundamentally different from the case that a net external
force is dragging the particle or stirring the fluid. Motivated by the
experimental realizations, the theoretical modelling of phoresis
relies on the assumption of ``low velocities'', which means a
stationary state ($\partial_t\equiv 0$), a small Mach number (the
flows are incompressible), a small Reynolds number (nonlinear,
convective effects are neglected in the Navier--Stokes equations), and
small Peclet numbers (one dismisses the advection of other relevant
fields like mass concentration, temperature, electric charge
distribution \dots). Under these conditions, the dynamic formulation
splits neatly into two different problems:
\begin{itemize}
\item The hydrodynamic problem for a
forced creep flow,
\begin{equation}
  \label{eq:creep}
  \left.
    \begin{array}[c]{c}
      \nabla\cdot\bu = 0 ,
      \\
      \\
      \eta \nabla^2 \bu - \nabla p = - \mathbf{f}
    \end{array}
  \right\}
  \qquad\Rightarrow\qquad
  \eta\nabla^2 (\nabla\times\bu) = - \nabla\times\mathbf{f} ,
\end{equation}
i.e., the incompressible, Stokes equation for the balance between the
fluid stresses and a force density field $\mathbf{f}(\br)$.

\item The problem that describes how the nonequilibrium state is
  established, which determines the kind of phoresis:
  diffusiophoresis, thermophoresis, electrophoresis, \dots This
  problem is decoupled from the flow because of the low Peclet number
  and can be solved on its own. For instance, in the case of
  diffusiophoresis in a solution with a single, neutral solute, the
  hypothesis of local equilibrium implies
\begin{equation}
  \label{eq:force}
  \mathbf{f} = - c \nabla \mu ,
\end{equation}
where $c(\br)$ is the solute concentration field, and $\mu[c]$ is the
associated chemical potential as a functional of $c(\br)$, i.e., the
equation of state. Mass balance then leads to the problem
\begin{equation}
  \label{eq:massbalance}
  \left.
    \begin{array}[c]{c}
      \displaystyle \frac{\partial c}{\partial t} = - \nabla\cdot\bj = 0
      \\
      \\
      \bj := \Gamma \mathbf{f}
    \end{array}
  \right\}
  \qquad\Rightarrow\qquad
  \nabla\cdot\left( \Gamma c \nabla\mu \right) = 0 , 
\end{equation}
where $\Gamma$ is the solute mobility that relates the force with the
solute current density $\bj(\br)$.
\end{itemize}

\subsection{The hydrodynamic problem}
\label{sec:hydrodynamic-problem}

The problem posed by \eqs{eq:creep} is common to the phoretic
phenomenology. It has to be complemented by boundary conditions:
\begin{itemize}

\item A vanishing flow at infinity, which sets the reference system
  (the ``laboratory frame'') with respect to which velocities are
  measured,
  \begin{equation}
    \label{eq:uinfty}
    \bu(\br)\to 0
    \qquad\mathrm{as}\qquad
    |\br|\to\infty .
  \end{equation}

\item A condition on the surface of the phoretic particle, usually a
  no-slip constraint,
  \begin{equation}
    \label{eq:noslip}
    \bu(\br) = \bV + \bOmega\times\br ,
    \qquad\mathrm{as}\qquad
    \br\in\textrm{particle's surface} ,
  \end{equation}
  where $\bV$ is the velocity of translation of the particle, and
  $\bOmega$ is the rotational velocity. 

\item These magnitudes are unknown, being fixed by the constraint of
  mechanical isolation: assuming that the force field $\mathbf{f}$
  vanishes with distance sufficiently fast, as corresponds to a local
  interaction between the particle and the fluid, the absence of
  external forces and torques means that the fluid does not transmit
  stresses across a surface $\mathcal{S}_\infty$ located at infinity,
  \begin{equation}
    \label{eq:mechisol}
    \oint_{\mathcal{S}_\infty}
    d\boldsymbol{\mathcal{S}}\cdot\mathsf{\Pi} = 0 ,
    \qquad
    \oint_{\mathcal{S}_\infty} \left(
      d\boldsymbol{\mathcal{S}}\cdot\mathsf{\Pi}
      \right)\times\br = 0 ,
    \end{equation}
    in terms of the fluid stress tensor (already simplified by the
    incompressibility constraint),
    \begin{equation}
      \label{eq:Pi}
      \mathsf{\Pi} = \eta \left[ \nabla\bu + (\nabla\bu)^\dagger
      \right] - \mathit{I} p .
    \end{equation}
  
\end{itemize}
If the purpose is only the determination of the phoretic velocities
$\bV$ and $\bOmega$, one can sidestep the computation of the whole
flow field $\bu(\br)$ by applying the Lorentz reciprocal theorem (see,
e.g., \rcites{HaBr73,KiKa91,Teub82,DPRD20}). In this manner one can
express the phoretic velocities explicitly as linear functionals of
$\nabla\times\mathbf{f}$ \cite{DoPo24b},
\begin{equation}
  \label{eq:Vgeneral}
  \bV = \int\limits_\mathrm{fluid} d^3\br\; \mathsf{M}^{(V)}(\br) \cdot
  \left[ \nabla \times \mathbf{f}(\br) \right] ,
\end{equation}
\begin{equation}
  \label{eq:Omgeneral}
  \bOmega = \int\limits_\mathrm{fluid} d^3\br\; \mathsf{M}^{(\Omega)}(\br) \cdot
  \left[ \nabla \times \mathbf{f}(\br) \right] ,
\end{equation}
where the tensor fields $\mathsf{M}^{(V)}(\br)$ and
$\mathsf{M}^{(\Omega)}(\br)$ are determined completely by the
geometrical shape of the particle. For example, a spherical particle
of radius $R$ gives (in dyadic notation)
\begin{equation}
  \label{eq:MVspher}
  \mathsf{M}^{(V)}(\br) = \frac{1}{6\pi \eta R} A(r) \;
  \underbrace{\left( \be_x \be_x + \be_y \be_y + \be_z \be_z
    \right)}_{\textrm{identity tensor } \mathsf{I}}
  \times\frac{\br}{r} ,
  \qquad
  A(r) := \frac{3 R}{4} \left[
    1 - \frac{2}{3} \frac {r}{R} - \frac{1}{3} \left(\frac{R}{r}\right)^2
  \right] ,
\end{equation}
\begin{equation}
  \label{eq:MOmspher}
  \mathsf{M}^{(\Omega)}(\br) = \frac{1}{2\pi\eta R^3} \Phi(r) \;
  \overbrace{\left( \be_x \be_x + \be_y \be_y + \be_z \be_z
    \right)}^{\textrm{identity tensor } \mathsf{I}} ,
  \qquad
  \Phi(r) := \frac{3 R^2}{2} \left[
    \frac{1}{3} \left(\frac{r}{R}\right)^2 + \frac{2}{3} \frac{R}{r} -
    1
  \right] .
\end{equation}

\subsection{Source of nonequilibrium}
\label{sec:source}

The force field $\mathbf{f}$ is obtained by solving a boundary value
problem in the fluid domain consisting of \eq{eq:massbalance} and of
boundary conditions: as for the latter, we focus on the case of
self-phoretic particles, i.e., we do not address the case of an
externally imposed gradient in concentration, being generated instead
only by a catalytic activity on the surface of the phoretic
particle. In such case, the boundary conditions read:
\begin{itemize}
\item An equilibrium, homogeneous state at infinity,
  \begin{equation}
    \label{eq:cinfty}
    c(\br)\to c_0 ,
    \;
    \mu(\br) \to \mu_0,
    \qquad\mathrm{as}\qquad
    |\br|\to\infty .
  \end{equation}
\item A nonvanishing solute flux at the surface of the particle,
  \begin{equation}
    \label{eq:activity}
    \bn\cdot\bj(\br) = \act(\br),
    \qquad\mathrm{as}\qquad
    \br\in\textrm{particle's surface} ,
  \end{equation}
  where $\bn$ is the unit vector normal to the surface, and the
  \emph{activity} $\act(\br)$ is the rate of the chemical reaction
  with which solute molecules are generated (or destroyed) at each
  surface point $\br$.
\end{itemize}
The activity pattern $\act(\br)$ is the source of a nonequilibrium
state, and thus of phoresis: upon setting $\act=0$, the
equilibrium state ($\nabla\mu=0 \; \Rightarrow\;\mathbf{f}=0$) is
recovered and no phoretic motion is induced ($\bV=0, \bOmega=0$ by
Eqs.~(\ref{eq:Vgeneral}, \ref{eq:Omgeneral})).

This boundary value problem is usually not amenable to an analytical
approach, unless simplifications are used. Although one can proceed
with a generic form of the equation of state $\mu[c]$, the relevant
features of the correlation--induced self-phoresis are captured with
the following simple model:
\begin{equation}
  \label{eq:mu}
  \mu = \mu_0 + k_B T \left[ c \ln \frac{c}{c_0} - \corr^2 \nabla^2 c \right].
\end{equation}
The first term is the ideal gas contribution, while the second
accounts for solute--solute interactions, quantified by a correlation
length $\corr$. One can then look for the mathematical solution as a
perturbative expansion in $\act$ (weak activity limit, not unlike what
is done in \sref{sec:weakW}), and linearize in the deviations
$\delta c(\br) = c(\br)-c_0$ and $\delta \mu(\br) = \mu(\br)-\mu_0$
from the homogeneous state (determined by the boundary condition at
infinity). In this manner, one arrives at the following problem for
the chemical potential:
\begin{equation}
  \label{eq:mubvp}
  \nabla^2\delta\mu = 0 ,
  \qquad
  \bn\cdot\nabla\delta\mu = - \frac{\act(\br)}{\Gamma c_0}
  \;\textrm{at the particle's surface},
  \qquad
  \delta\mu=0
  \;\textrm{at infinity} ,
\end{equation}
from whose solution one obtains the concentration profile by
linearizing the equation of state~(\ref{eq:mu}):
\begin{equation}
  \label{eq:cbvp}
  \corr^2 \nabla^2 \delta c - \delta c = - \frac{\delta\mu(\br)}{k_B T} ,
  \qquad
  \bn\cdot\nabla\delta c = 0
  \;\textrm{at the particle's surface},
  \qquad
  \delta c=0
  \;\textrm{at infinity} .
\end{equation}
Two relevant features follow immediately from these equations: first,
since both $\delta\mu$ and $\delta c$ are linear in the activity, the
phoretic velocities will be quadratic in $\act$. And second, in the
absence of correlations ($\corr=0$), no flow or phoretic motion will
be induced because $\nabla\delta c$ and $\nabla\delta\mu$ will be
parallel, which justifies the denomination ``correlation--induced''
phoresis coined in \rcite{DPRD20}.

\subsection{Granular  self-phoresis}
\label{sec:selfph}

\begin{table}[t]
  \centering
\begin{tabular}[c]{c || c | c}
  &
    \begin{tabular}[c]{c}
      \textbf{Active particle $\act(\br)$}
      \\
      \textbf{in a fluid solution}
    \end{tabular}
  &
    \begin{tabular}[c]{c}
      \textbf{Intruder $\pot(\br)$ in}
      \\
      \textbf{a granular bath}
    \end{tabular}
  \\
  \hline\hline
  \\
  \begin{tabular}[c]{c}
    HYDRODYNAMIC
    \\
    \\
    PROBLEM
  \end{tabular}
  & $\quad$
    \begin{tabular}[c]{c}
      $\nabla\cdot\bu=0$
      \\
      \\
      $\eta\nabla^2(\nabla\times\bu) = \nabla\delta c \times
      \nabla\delta \mu$
      \\
      \\
      $\bu(\br\in\mathrm{particle})=\bV+\bOmega\times\br$
      \\
      \\
      $\bu(r\to\infty) = 0$
      \\
      \\
      $\displaystyle
      \bV = \int\limits_\mathrm{fluid} d^3\br\; \mathsf{M}^{(V)}(\br) \cdot
      \left[ \nabla\delta \mu(\br) \times \nabla\delta c(\br) \right]$
      \\
      \\
      $\displaystyle
      \bOmega = \int\limits_\mathrm{fluid} d^3\br\;
      \mathsf{M}^{(\Omega)}(\br)
      \cdot \left[ \nabla\delta \mu(\br) \times \nabla\delta c(\br) \right]$
    \end{tabular}
  $\quad$
  & $\quad$
    \begin{tabular}[c]{c}
      $\nabla\cdot\bu=0$
      \\
      \\
      $\eta\nabla^2(\nabla\times\bu) = \nabla\rho \times
      \nabla\pot$
      \\
      \\
      $\bu(\br\in\mathrm{intruder})=\bV+\bOmega\times\br$
      \\
      \\
      $\bu(r\to\infty) = 0$
      \\
      \\
      $\displaystyle
      \bV = \int\limits_\mathrm{fluid} d^3\br\; \mathsf{M}^{(V)}(\br) \cdot
      \left[ \nabla\pot (\br) \times \nabla \rho(\br) \right]$
      \\
      \\
      $\displaystyle
      \bOmega = \int\limits_\mathrm{fluid} d^3\br\;
      \mathsf{M}^{(\Omega)}(\br)
      \cdot \left[ \nabla\pot (\br) \times \nabla\rho(\br) \right]$
    \end{tabular}
  $\quad$
  \\
  \\
  \hline
  \\
  \\
  \begin{tabular}[c]{c}
    SOURCE OF
    \\
    \\
    NONEQUILIBRIUM
    \\
    \\
    PROBLEM
  \end{tabular}
  & $\quad$
    \begin{tabular}[t]{c}
      $\displaystyle \corr^2\nabla^2\delta c - \delta c = -
      \frac{\delta\mu(\br)}{k_B T}$
      \\
      \\
      $\bn\cdot\nabla\delta c(\br\in\mathrm{particle})=0$
      \\
      \\
      $\delta c(r\to\infty)=0$
      \\
      \\
      $\nabla^2\delta\mu=0$
      \\
      \\
      $\displaystyle \bn\cdot\nabla\delta
      \mu(\br\in\mathrm{particle})=-\frac{\act(\br)}{\Gamma c_0}$
      \\
      \\
      $\delta \mu(r\to\infty)=0$
    \end{tabular}
  $\quad$
  & $\quad$
    \begin{tabular}[t]{c}
      $\displaystyle \corr^2\nabla^2\rho - \rho = \gamma \pot(\br)$
      \\
      \\
      $\bn\cdot\nabla\rho(\br\in\mathrm{intruder})=0$
      \\
      \\
      $\rho(r\to\infty)=0$
      \\
      \\
    \end{tabular}
  $\quad$
\end{tabular}
  \caption{Comparison between the mathematical models for
    self-phoresis of an active particle in a fluid solution and an
    intruder in a granular bath}
  \label{tab:selfphoresis}
\end{table}

The analogy between the models of catalytic--activity and
granular--bath self-phoresis should now be straightforward. First,
one notices that the potential $\pot$ created by a moving intruder
would change in time because it depends on the intruder's position and
orientation. However, in the weak--field limit leading to the
perturbative equations~(\ref{eq:plin}--\ref{eq:flowlin}), the
velocities of translation and rotation will be found quadratic in
$\pot$, see table~\ref{tab:selfphoresis}, so that the time--dependence
of $\pot$ brought about by the intruder's motion will be a subleading
correction. Effectively, the weak--field approximation is equivalent
to the ``low velocity'' approximation discussed above, and it allows
one to address the influence of the potential $\pot$ to leading order
as if the intruder were instantaneously at rest.

Under these conditions, one can easily recognize the parallelism,
concerning the hydrodynamic problem, between the flow \eqs{eq:creep}
and Eqs.~(\ref{eq:divu}, \ref{eq:flowlin}) derived for the granular
fluid with the identification
$\nabla\times\mathbf{f} = \nabla\pot\times\nabla\rho$; when compared
with \eq{eq:force}, the most relevant feature is that the flow is
driven by a gradient misalignment in both cases. As for the problem
on the source of nonequilibrium, the correspondence with
\eq{eq:massbalance} is not obvious; but it becomes manifest with the
perturbative approximation, see \eqs{eq:cbvp}, when \eq{eq:heatlin} is
complemented by the boundary conditions that there is no heat flow
stemming either from the surface of the intruder in the granular bath
or from distant sources:
\begin{equation}
  \label{eq:rhobvp}
  \corr^2 \nabla^2 \rho - \rho = \gamma \pot ,
  \qquad
  \bn\cdot\nabla\rho = 0
  \;\textrm{at the intruder's surface},
  \qquad
  \rho=0
  \;\textrm{at infinity} .
\end{equation}
The parallelism in the mathematical models for the two phenomena of
self--phoresis is summarized in table~\ref{tab:selfphoresis}. It is
clear that one can borrow the results from the correlation--induced
self--phoresis of active particles to address the self-phoretic motion
of an intruder that acts on the granular bath through the potential
$\pot$. The granular scenario is actually somewhat simpler because the
potential $\pot$ is given, to be compared with the need to obtain
first $\delta\mu$ from a given activity pattern through an additional
boundary value problem. We here summarize the most relevant features
of the intruder's phoretic motion that follow from the mathematical
model. Since $\rho$ depends linearly on $\pot$, the phoretic
velocities will depend quadratically on the intruder's potential
$\pot(\br)$. One can also derive the asymptotic behaviors on the
length scale $\corr$. In the limit $\corr\to\infty$, one can drop the
screening term and approximate the Helmholtz equation for $\rho$ by a
Poisson equation, $\corr^2 \nabla^2 \rho \approx \gamma \pot$, so that
the length appears asymptotically only as a scale factor,
$\rho\sim\xi^{-2}$. In this limit, the length $\corr$ does not
contribute to the convergence of the integrals in
Eqs.~(\ref{eq:Vgeneral}, \ref{eq:Omgeneral}) because the integrands
already decay sufficiently rapidly at infinity through the tensorial
kernels $\mathsf{M}$, the potential $\pot$, and the field $\rho$, and
consequently $\bV, \bOmega \sim \xi^{-2}$. To address the opposite
limit ($\corr\to 0$), one introduces the variable
$\hat{\rho}:=\rho+\gamma\pot$, so that the flow source takes the form
$\nabla\hat{\rho}\times\nabla\pot$ and \eqs{eq:rhobvp} are transformed
into the following alternative boundary value problem:
\begin{equation}
  \label{eq:rhohatbvp}
  \corr^2 \nabla^2 \hat{\rho} - \hat{\rho} = \corr^2 \gamma \nabla^2 \pot ,
  \qquad
  \bn\cdot\nabla\hat{\rho} = \gamma \bn\cdot\nabla\pot
  \;\textrm{at the intruder's surface},
  \qquad
  \hat{\rho}=0
  \;\textrm{at infinity} .
\end{equation}
When $\corr\to 0$, the field $\hat{\rho}$ is fully screened and can be
approximated in the fluid bulk by a completely local relationship,
$\hat{\rho} \approx \corr^2 \gamma \nabla^2 \pot$; the boundary
condition at the surface of the particle can be ignored because its
effect is restricted to a thin layer of thickness $\sim\xi$ next to
the intruder's surface which gives a subdominant contribution to the
integrals~(\ref{eq:Vgeneral}, \ref{eq:Omgeneral}) because the
tensorial kernels $\mathsf{M}$ vanish\footnote{Specifically,
  $\mathsf{M}\sim z^2$ with the separation $z\to 0$ to the surface, so
  that this yields a contribution $\sim\xi^3$ to the phoretic
  velocities.} at the surface due to the no-slip boundary condition
\cite{DoPo24b}. As a consequence, one expects
$\bV, \bOmega \sim \xi^{2}$ in this limit.

Further, specific insight can be derived from the particular case of a
spherical intruder. By introducing spherical coordinates
$\{r,\theta,\varphi\}$ with origin at the intruder's center, one can
express the external potential and the heat potential as expansions
in spherical harmonics,
\begin{equation}
  \label{eq:expansionlm}
  \pot (\br) = \sum_{\ell m} w_{\ell m}(r) \, Y_{\ell m}
  (\theta,\varphi) ,
  \qquad
  \rho (\br) = \sum_{\ell m} \rho_{\ell m}(r) \, Y_{\ell m}
  (\theta,\varphi) .  
\end{equation}
One now proceeds by computing first the heat potential and the
phoretic velocities successively:
\begin{enumerate}
\item The boundary value problem~(\ref{eq:rhobvp}) then becomes a problem
that determines the coefficients $\rho_{\ell m}(r)$:
\begin{equation}
  \label{eq:rhobvp2}
 \frac{d^2 \rho_{\ell m}}{d r^2} + \frac{2}{r} \frac{d\rho_{\ell m}}{d
   r} - \left[ \frac{\ell (\ell+1)}{r^2} + \frac{1}{\corr^2} \right]
 \rho_{\ell m} =
 \frac{\gamma}{\corr^2} w_{\ell m}(r)  ,
 \qquad
 \frac{d\rho_{\ell m}}{dr}(R)=0 ,
 \qquad
 \rho(r\to\infty)=0 .
\end{equation}
The solution is
\begin{equation}
  \label{eq:rholm}
  \rho_{\ell m} (r) = \int_R^\infty dr'\; 
  \mathcal{D}_\ell (r,r') \, w_{\ell m}(r') ,
\end{equation}
where $\mathcal{D}_\ell(r,r')$ is Green's function for the boundary
value problem~(\ref{eq:rhobvp2}) and can be expressed in terms of the
modified Bessel functions of order $\ell+1/2$, see the Supplemental
Material in \rcite{DPRD20}.

\item The translational velocity~(\ref{eq:Vgeneral}) with the
kernel~(\ref{eq:MVspher}) reads 
\begin{equation}
  \label{eq:Vspher}
  \bV = \frac{1}{6\pi\eta R} \int_R^\infty dr \; r^2 A(r)
  \int_0^\pi d\theta\;\sin\theta \int_0^{2\pi} d\varphi\;
  \be_r\times\left(\nabla\pot\times\nabla\rho\right) ,
\end{equation}
after the replacement $\nabla\times\mathbf{f} =
\nabla\pot\times\nabla\rho$. One splits the nabla operator into radial
and tangential components as
\begin{equation}
  \label{eq:nablaspher}
  \nabla = \be_r \frac{\partial}{\partial r}+ \frac{1}{r} \nabla_\parallel ,
  \qquad
  \nabla_\parallel = \be_\theta \frac{\partial}{\partial \theta}
  + \frac{\be_\varphi}{\sin\theta} \frac{\partial}{\partial \varphi} ,
\end{equation}
so that
\begin{equation}
  \label{eq:curlspher}
  \be_r\times\left(\nabla\pot\times\nabla\rho\right) =
  \be_r\times\left[
    \frac{1}{r^2} \nabla_\parallel\pot\times\nabla_\parallel\rho
    + \frac{\be_r}{r}\times \left(
      \frac{\partial \pot}{\partial r} \, \nabla_\parallel\rho
      - \frac{\partial \rho}{\partial r} \, \nabla_\parallel\pot
    \right)
  \right]
  = \frac{\partial \rho}{\partial r} \, \nabla_\parallel\pot
  - \frac{\partial \pot}{\partial r} \, \nabla_\parallel\rho .
\end{equation}
By inserting the expansions~(\ref{eq:expansionlm}) and reordering, one
can finally write
\begin{equation}
  \label{eq:Vspher3}
  \bV = \sum_{\ell m} \sum_{\ell' m'} h^\parallel_{\ell m; \ell' m'} \,
  \bG^\parallel_{\ell m; \ell' m'} ,
\end{equation}
where the radial integral leads to the coefficients
\begin{equation}
  \label{eq:hV}
  h^\parallel_{\ell m; \ell' m'} :=
  -\frac{\ell'+1}{6\pi\eta R}
  \int_R^\infty dr \; r \, A(r) \left[
    \frac{d \rho_{\ell m}}{d r} \, w_{\ell' m'}(r)
    - \frac{d w_{\ell m}}{d r} \, \rho_{\ell' m'}(r)
  \right] ,
\end{equation}
which are quadratic functionals of the functions $w_{\ell m}(r)$ in
view of the relationship~(\ref{eq:rholm}), while the angular integrals
are collected in the following vectors defined in \rcite{DPRD20}:
\begin{equation}
  \label{eq:Gpar}
  \bG^\parallel_{\ell m; \ell' m'} := - \frac{1}{\ell'+1}
  \int_0^\pi d\theta\;\sin\theta \int_0^{2\pi} d\varphi\;
  Y_{\ell m}(\theta,\varphi)
  \, \nabla_\parallel Y_{\ell' m'}(\theta,\varphi) ,
\end{equation}
These vectors can be expressed in terms of the Wigner 3j
symbols. Despite appearances, \eq{eq:Vspher3} is not a double
summation because the vectors $\bG^\parallel$ vanish for almost all
the index combinations and the expression reduces to a single
summation; more precisely, the following ``selection rules'' hold:
\begin{equation}
  \label{eq:selectionPar}
  \bG^{\parallel}_{\ell m; \ell' m'} = 0
  \quad\textrm{ if }\quad
  \left\{
    \begin{array}[c]{l}
      \ell-\ell' \neq \pm 1 , \; \mathrm{or}
      \\
      m+m' \neq 0, \pm 1 ,
    \end{array}
  \right.
\end{equation}
which imply that translational phoresis necessarily requires a
potential $\pot(\br)$ which has nonvanishing, neighbouring multipolar
moments.

\item One proceeds similarly with the phoretic angular velocity
$\bOmega$, although in this case it is mathematically simpler if one
first integrates by parts in \eq{eq:Omgeneral} and applies that
$\Phi(R)=0$:
\begin{eqnarray}
  \label{eq:Omspher}
  \bOmega
  & =
  & \frac{1}{2\pi\eta R^3} \int\limits_{r>R} d^3\br\;
  \Phi(r) \overbrace{\nabla\pot (\br) \times
    \nabla\rho(\br)}^{\nabla\times(\pot\nabla\rho)}
    \nonumber
  \\
  & =
  & \frac{1}{2\pi\eta R^3} \left\{
    \oint_{r=R} d\boldsymbol{\mathcal{S}}\times[\nabla\rho(\br)]\; \pot(\br) \Phi(r)
    - \int\limits_{r>R} d^3\br\;
    \overbrace{[\nabla\Phi(r)]}^{\be_r \Phi'(r)} \times
    [\nabla\rho(\br)] \pot(\br)
    \right\}
    \nonumber \\
  & =
  & - \frac{1}{2\pi\eta R^3} \int_R^\infty dr \; r \, \Phi'(r)
  \int_0^\pi d\theta\;\sin\theta \int_0^{2\pi} d\varphi\;
    \pot(\br) \, \be_r\times\nabla_\parallel\rho(\br).
\end{eqnarray}
Inserting again the the expansions~(\ref{eq:expansionlm}) and
reordering, one gets
\begin{equation}
  \label{eq:Omspher3}
  \bOmega = \sum_{\ell m} \sum_{\ell' m'} h^\tau_{\ell m; \ell' m'} \,
  \bG^\tau_{\ell m; \ell' m'} ,
\end{equation}
in terms of the coefficients
\begin{equation}
  \label{eq:hOm}
  h^\tau_{\ell m; \ell' m'} := \frac{\ell'+1}{2\pi\eta R^3}
  \int_R^\infty dr \; r \, \Phi'(r) \, w_{\ell m}(r)\, \rho_{\ell' m'}(r) ,
\end{equation}
and the vectors \cite{DPRD20}
\begin{equation}
  \label{eq:Gtau}
  \bG^\tau_{\ell m; \ell' m'} := - \frac{1}{\ell'+1}
  \int_0^\pi d\theta\;\sin\theta \int_0^{2\pi} d\varphi\;
  Y_{\ell m}(\theta,\varphi)
  \, \be_r (\theta,\varphi) \times\nabla_\parallel Y_{\ell' m'}(\theta,\varphi) .
\end{equation}
Again, these vectors can be expressed in terms of the Wigner 3j
symbols and obey a set of ``selection rules'':
\begin{equation}
  \label{eq:selectionTau}
  \bG^{\tau}_{\ell m; \ell' m'} = 0
  \quad\textrm{ if }\quad
  \left\{
    \begin{array}[c]{l}
      \ell-\ell' \neq 0 , \; \mathrm{or}
      \\
      m=m' , \; \mathrm{or}
      \\
      m+m' \neq 0, \pm 1 .
    \end{array}
  \right.
\end{equation}
The phoretic angular velocity already represents a main departure from
the scenario of an active particle because, in the latter case,
$\bOmega=0$ always \cite{DPRD20}. The reason is that the activity
pattern $\act$ defined in \eq{eq:activity} is a function fully
localized at the surface of the particle: an expansion
like~(\ref{eq:expansionlm}) yields $r$--independent numbers
$a_{\ell m}$ that factor out of the radial integrals~(\ref{eq:rholm})
and~(\ref{eq:hOm}), in which case one can proof that the sums over $m$
and $m'$ in \eq{eq:Omspher3} vanish identically. This reflects that
chirality is not broken in this case and no preferred direction of
rotation emerges. On the contrary, this is not the case for the
intruder because the potential $\pot(\br)$ also acts in the bulk of
the granular bath, and the appearance of the functions $w_{\ell m}(r)$
inside the radial integrals invalidate the proof: the chirality is
then broken by the different behavior under rotations (encoded by the
indices $m$, $m'$) in the two fields $w_{\ell m}(r)$ and
$\rho_{\ell m}(r)$ that appear in \eq{eq:hOm}.

\end{enumerate}

As an application of these results, we consider the simplest example
of a potential that leads to self-phoresis,
\begin{equation}
  \label{eq:partW}
  \pot(\br) = W(r) \left( 1 + p \cos \theta \right) ,
  \qquad
  W(r) := \pot_0 \mathrm{e}^{-r/R} .
\end{equation}
($\pot_0$ and $p$ are given constants, which in the final
result~(\ref{eq:partV}) will just appear as a prefactor and can be
thus set equal to unity without loss of generality; the radial
dependence is chosen to facilitate the analytical calculations). This
potential contains only a monopole and a dipole, which is the simplest
description of a bifaced or Janus particle:
\begin{equation}
  \label{eq:partw}
  w_{00}(r) = \sqrt{4\pi} \, W(r) ,
  \qquad\qquad
  w_{10}(r) = \sqrt{\frac{4\pi}{3}} \, p\, W(r) ,
  \qquad\qquad
  w_{\ell m}(r) = 0 \textrm{ otherwise}.
\end{equation}
 (Notice that the orientation of the $z$ axis can always be chosen such
that $w_{1,\pm 1}=0$ without loss of generality.) The
summation~(\ref{eq:Vspher3}) reduces to a single term,
\begin{equation}
  \label{eq:partV}
  \bV =h^\parallel_{0 0; 1 0} \bG^\parallel_{0 0; 1 0} = -
  \frac{h^\parallel_{0 0; 1 0}\, \be_z}{\sqrt{3}} ,
  \qquad
    h^\parallel_{0 0; 1 0} :=
  -\frac{1}{3\pi\eta R}
  \int_R^\infty dr \; r \, A(r) \left[
    \frac{d \rho_{0 0}}{d r} \, w_{1 0}(r)
    - \frac{d w_{0 0}}{d r} \, \rho_{1 0}(r)
  \right] .
\end{equation}
Although the coefficients $\rho_{00}(r), \rho_{10}(r)$ and the
integral $h^\parallel_{0 0; 1 0}$ can be evaluated analytically, we
omit the resulting lengthy expression; it is more illuminating to show
\fref{fig:selfphoresis} instead: the phoretic velocity has a maximum
as a function of $\corr$ and exhibits the theoretically predicted
asymptotic behaviors. The angular velocity $\bOmega$ vanishes in this
example due to the selection rules~(\ref{eq:selectionTau}); a
potential $\pot$ with at least quadrupolar components is needed to
have a nonvanishing angular velocity.

\begin{figure}[t]
  \centering
  \hfill\includegraphics[width=.45\textwidth]{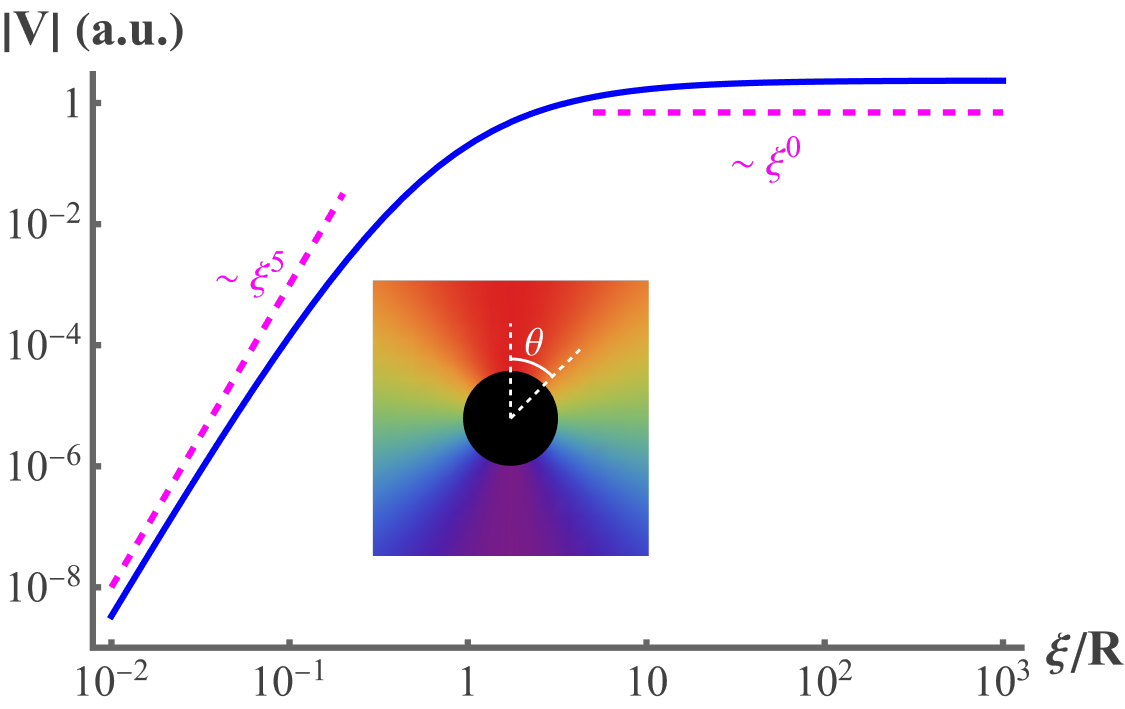}\hspace*{\fill}
  \caption{Correlation--induced self-phoretic translation velocity
    $|\bV|$ (in arbitrary units) as a function of the ratio $\corr/R$,
    for the case of a spherical active particle of radius $R$ with the
    activity pattern~(\ref{eq:partA}), whose angular dependence on the
    particle's surface is shown in the inset as a heat map. The
    expected asymptotic behaviors are also shown.}
  \label{fig:activeSelfphoresis}
\end{figure}

For comparison, \fref{fig:activeSelfphoresis} shows the same plot
(phoretic velocity as a function of the correlation length $\corr$) in
the case of correlation--induced self-phoresis of an active particle,
when taking an activity pattern of the same kind as before (monopole +
dipole, $\act_0$ and $p$ are again given constants):
\begin{equation}
  \label{eq:partA}
  \act(\br=R\be_r) =  \act_0 \left( 1 + p \cos \theta \right) 
  \qquad\Rightarrow\qquad
  \left\{
    \begin{array}[c]{l}
      a_{00} = \sqrt{4\pi} \, \act_0 ,
      \\
      \\
      a_{10} = \displaystyle \sqrt{\frac{4\pi}{3}} \, p \, \act_0 ,
      \\
      \\
      a_{\ell m} = 0 \textrm{ otherwise},
    \end{array}
  \right.
\end{equation}
where the coefficients $a_{\ell m}$ are defined by an expansion like
in \eq{eq:expansionlm} but for $\act$.  The phoretic velocity is given
again by \eq{eq:partV} but, in view of table~(\ref{tab:selfphoresis}),
with a different radial integral:
\begin{equation}
  \label{eq:hVact}
  \bV =h^\parallel_{0 0; 1 0} \bG^\parallel_{0 0; 1 0} = -
  \frac{h^\parallel_{0 0; 1 0}\, \be_z}{\sqrt{3}} ,
  \qquad
  h^\parallel_{0 0; 1 0} :=
  -\frac{1}{3\pi\eta R}
  \int_R^\infty dr \; r \, A(r) \left[
    \frac{d c_{0 0}}{d r} \, \mu_{1 0}(r)
    - \frac{d \mu_{0 0}}{d r} \, c_{1 0}(r)
  \right] ,
\end{equation}
whereby the expansion coefficients are obtained by solving
Eqs.~(\ref{eq:mubvp}) and~(\ref{eq:cbvp}), respectively, giving
\begin{equation}
  \label{eq:mulm}
  \mu_{00}(r) = \frac{a_{0 0} R^2}{\Gamma c_0 r} ,
  \qquad
  \mu_{10}(r) = \frac{a_{1 0} R^3}{2 \Gamma c_0 r^2} ,
\end{equation}
and a somewhat more involved expressions for $c_{00}$ and $c_{10}$
(see the Supplemental Material in \rcite{DPRD20}). The dependence of
$\bV$ on $\corr$ exhibits qualitative differences between
\fref{fig:selfphoresis} and \fref{fig:activeSelfphoresis}, which can
be traced back to the fact that, unlike the generic case of
$\pot(\br)$, the solute chemical potential $\delta\mu(\br)$ is a
harmonic function, see \eq{eq:mubvp}.

\section{Molecular Dynamics Simulations}
\label{sec:simul}

We have simulated the dynamics of a collection of inelastic particles,
using the energy injection model proposed in \rcite{BRS13} and
incorporating an external force potential. The latter is approximated
by a discretized version of \(\pot\), allowing the implementation of a
precise event--driven algorithm \cite{PoSc05}. Particles are modelled
as hard disks of unit mass and unit diameter that move within a
rectangular box of width \(L_x\) and height \(L_y\), with periodic
boundary conditions. In addition, a grid of square cells of unit
length overlaps the box and serves the purpose of discretizing the
potential: inside a given cell with center point \((x_0,y_0)\), the
continuous potential \(\pot(x,y)\) is approximated by
\(\tilde\pot(x,y)=\pot(x_0,y_0)\), so that \(\tilde\pot\) becomes a
piecewise constant function.

$\;$

\noindent \textbf{Event--driven algorithm:} Before considering
collisions, the particles move freely within a cell. Therefore, the
dynamics is driven by two events:
\begin{itemize}
\item[(a)] either the center of a disk reaches the edge of a cell, 
\item[(b)] or two disks touch.
\end{itemize}
Assuming that only one event can take place at a given time at
most\footnote{With unlimited precision, this is true except for a set
  of measure zero. In general, it is a very good assumption if
  pathological initial conditions are removed.}, the implementation of
an event--driven algorithm is straightforward. It relays on two
elementary steps:
\begin{itemize}
\item[(i)] For a given physical configuration without overlapping
  particles, the time step till the next event is calculated. This
  requires computing, for each particle, the time step till the next
  event of kind (a) and kind (b).
\item[(ii)] The next event is identified by identifying the minimal
  time step (which, by assumption, is unique) out of the list of all
  the time steps computed in (i). All particles update their
  positions using their current velocities and this minimal time
  step. The velocities of the particles involved in the event are also
  updated according to the kind of event:
  \begin{itemize}
  \item[(a)] A particle has reached an edge: if the velocity component
    normal to the edge is sufficiently large to overcome the potential
    difference between the two neighbouring cells, the particle
    crosses to the corresponding cell and its normal velocity is
    reduced accordingly. Otherwise, the particle bounces off. In both
    cases, the mechanical energy is conserved.
  \item[(b)] Two particles collide: if their precollisional velocities
    are \(\boldsymbol v_1\) and \(\boldsymbol v_2\), respectively,
    they acquire postcollisional velocities given as
    \begin{eqnarray}
      \boldsymbol v_1^*
      & =
      & \displaystyle \boldsymbol v_1-\frac{1+\alpha}{2} [ (\boldsymbol
        v_1-\boldsymbol v_2)\cdot \hat\sigma] \hat\sigma-\hat
        \sigma\Delta,
      \\
      \boldsymbol v_2^*
      & =
      & \displaystyle \boldsymbol v_1+\frac{1+\alpha}{2} [ (\boldsymbol
        v_1-\boldsymbol v_2)\cdot \hat\sigma] \hat\sigma+\hat
        \sigma\Delta.
    \end{eqnarray}
    Here, the parameter \(\alpha\in[0,1)\) is the coefficient of
    normal restitution accounting for the inelastic character of the
    collisions, \(\hat \sigma\) is a unit vector pointing from the
    center of particle \(1\) to that of particle \(2\), and
    \(\Delta>0\) models the bulk injection of a fixed amount of energy
    by an unspecified external source. This collision rule conserves
    linear momentum and angular momentum (with respect to the
    collision contact point), but not energy in general.
  \end{itemize}
\end{itemize}
The actual implementation of this event--driven algorithm has been
done efficiently by eliminating unnecessary calculations of event
times steps and updates \cite{Lu91}.

$\;$

\noindent \textbf{Simulation:} It consists of the following steps:
\begin{enumerate}
\item Setting the values of the parameters:
  \begin{itemize}
  \item[-] Number of particles \(N\) and system size \(L_x\), \(L_y\):
    variable, but such that the particle density \(N/(L_x\,L_y)\) is
    of the order of \(0.028\) (dilute regime).
  \item[-] The temperature of the initial condition is set to
    \(1\). This, together with the unit mass and diameter of the
    disks, defines the unit of time.
  \item[-] Coefficient of normal restitution \(\alpha=0.85\) and
    energy injection parameter \(\Delta=0.1643\). With this choice,
    the dissipation is moderate, and the final overall asymptotic
    temperature of the system remains close to \(1\).
  \item[-] External potential given by \eq{eq:Wpart} with a
    coefficient \(\pot_0=2\). This value emerges from a compromise: on
    the one hand, it ensures that the particle distribution does not
    deviate much from homogeneity. But, on the other hand, it confines
    the particles such that the center of mass of the system does not
    wander significantly during the total time of simulation,
    executing instead a Brownian motion around the center of the box.
  \end{itemize}
\item Setting the initial state: The particles are distributed
  homogeneously inside the box and with a Gaussian distributed
  velocity. The total linear momentum and the angular momentum with
  respect to the box center are explicitly set to zero.
\item Transient evolution to a steady state: The event--driven
  algorithm is run \(10^7\) time units. This results in roughly
  \(10^5\) collisions per particle, which ensures that the system has
  reached a steady state.
\item Measuring observables: The algorithm is run for another time window of
  \(10^7\) units, collecting measurements every time unit. These
  include the number of particles, the total linear momentum, and the
  total energy of each cell.
\item Averaging: The steps (2)-(4) constitute a single
  realization. The hydrodynamic fields (velocity, density,\dots),
  discretized on the grid, are obtained by averaging over \(10^2\)
  different realizations.
\end{enumerate}

\newpage